\documentclass[structabstract,english,latin9]{aa}

\usepackage{graphicx}
\usepackage{epstopdf}
\usepackage{epsfig}
\usepackage{color}
\usepackage{multirow}
\usepackage{rotating}
\usepackage{lscape}
\usepackage{longtable}
\usepackage{lipsum}
\usepackage{txfonts}
\usepackage{amssymb}
\usepackage{pdflscape}
\usepackage{lipsum, babel}
\usepackage{multicol}
\usepackage{subfig}

\providecommand{\tabularnewline}{\\}

\titlerunning {Radio AGN dichotomy with quasar clustering}
\authorrunning{Retana-Montenegro et al. 2016}

\makeatother

\usepackage{babel}
\begin{document}

\title{}

\title{Probing the Radio Loud/Quiet AGN dichotomy with quasar clustering}

\author{E. Retana-Montenegro\inst{1}\and  H. J. A. R\"{o}ttgering\inst{1}
}

\offprints{E. Retana-Montenegro}

\institute{Leiden Observatory, Leiden University, P.O. Box 9513, 2300 RA, Leiden,
The Netherlands\\
\email{eretana@strw.leidenuniv.nl}\\
}

\date{Received September xx, xxxx; accepted March xx, xxxx}

\abstract{We investigate the clustering properties of 45441 radio-quiet quasars
(RQQs) and 3493 radio-loud quasars (RLQs) drawn from a joint use of
the Sloan Digital Sky Survey (SDSS) and Faint Images of the Radio
Sky at 20 cm (FIRST) surveys in the range $0.3<z<2.3$. This large
spectroscopic quasar sample allow us to investigate the clustering
signal dependence on radio-loudness and black hole (BH) virial mass.
We find that RLQs are clustered more strongly than RQQs in all the
redshift bins considered. We find a real-space correlation length
of $r_{0}=6.59_{-0.24}^{+0.33}\,h^{-1}\,\textrm{Mpc}$ and $r_{0}=10.95_{-1.58}^{+1.22}\,h^{-1}\,\textrm{Mpc}$
{\normalsize{}for} RQQs and RLQs, respectively, for the full redshift
range. This implies that RLQs are found in more massive host haloes
than RQQs in our samples, with mean host halo masses of $\sim4.9\times10^{13}\,h^{-1}\,M_{\odot}$
and $\sim1.9\times10^{12}\,h^{-1}\,M_{\odot}$, respectively. Comparison
with clustering studies of different radio source samples indicates
that this mass scale of $\gtrsim1\times10^{13}\,h^{-1}\,M_{\odot}$
is characteristic for the bright radio-population, which corresponds
to the typical mass of galaxy groups and galaxy clusters. The similarity
we find in correlation lengths and host halo masses for RLQs, radio
galaxies and flat-spectrum radio quasars agrees with orientation-driven
unification models. Additionally, the clustering signal shows a dependence
on black hole (BH) mass, with the quasars powered by the most massive
BHs clustering more strongly than quasars having less massive BHs.
We suggest that the current virial BH mass estimates may be a valid
BH proxies for studying quasar clustering. We compare our results
to a previous theoretical model that assumes that quasar activity
is driven by cold accretion via mergers of gas-rich galaxies. While
the model can explain the bias and halo masses for RQQs, it cannot
reproduce the higher bias and host halo masses for RLQs. We argue
that other BH properties such as BH spin, environment, magnetic field
configuration, and accretion physics must be considered to fully understand
the origin of radio-emission in quasars and its relation to the higher
clustering.}

\maketitle

\section{Introduction}

Quasars are luminous active galactic nuclei (AGN) powered by supermassive
black holes (SMBHs) \citep{1964ApJ...140..796S,1969Natur.223..690L}.
The role of AGN activity in galaxy formation and evolution processes
is still not well understood. Evidence for a co-evolution scenario
is provided by the empirical relationship between the host galaxy
velocity dispersion and the mass of their central black holes (BHs)
\citep{2000ApJ...539L...9F,2000ApJ...539L..13G}. At low-z, the analysis
of stars and gas dynamics in the nucleus of nearby galaxies \citep{2006ApJ...646..754D,2006AA...460..439D,2007AA...469..405P,2008AA...479..355D,2009ApJ...693..946S,2013ApJ...770...86W}
and the reverberation mapping technique \citep{1988PASP..100...18P,2004ApJ...613..682P,2012MNRAS.426..416D,2012ApJ...755...60G}
have found that the most massive galaxies harbour the most massive
BHs. At high-z, virial BH mass ($M_{\textrm{BH}}$) estimations based
on single-epoch spectra employing empirical scaling relations (e.g.
\citealp{2000ApJ...533..631K,2004MNRAS.352.1390M,2008ApJ...680..169S})
suggest that SMBHs with masses $>10^{9}\,M_{\odot}$ were already
in place at $z\gtrsim5$ \citep{2003ApJ...587L..15W,2007AJ....134.1150J,2011Natur.474..616M,2014arXiv1410.2689Y}.

Because of their high-luminosity, quasars are excellent tracers of
the large-scale structure up to $z\sim6$. Recent large optical surveys
using wide field integral spectrographs, such as the Sloan Digital
Sky Survey (SDSS, \citealp{2000AJ....120.1579Y}) and the 2dF QSO
Redshift Survey (2QZ, \citealp{2004MNRAS.349.1397C}) have revealed
thousands of previously unknown quasars. These newly detected quasars
can be used to construct large statistical samples to study quasar
clustering in detail across cosmic time. Several authors have found
that quasars have correlation lengths of $r_{0}=5\,h^{-1}-8.5\,h^{-1}\,\textrm{Mpc}$
at $0.8<z<2.0$, indicating that they reside in massive dark matter
haloes (DMH) with masses of $\sim10^{12}-10^{13}\,M_{\odot}$ (e.g.
\citealp{2004MNRAS.355.1010P,2006ApJ...638..622M,2008MNRAS.383..565D,2009ApJ...697.1634R,2009ApJ...697.1656S}). 

Such clustering measurements provide a means to probe the outcome
of any cosmological galaxy formation model \citep{2005Natur.435..629S,2008ApJS..175..356H},
to understand how SMBH growth takes place \citep{2005Natur.433..604D,2009MNRAS.396..423B,2010MNRAS.406.1959S},
to define the quasar host galaxies characteristic masses \citep{2010ApJ...718..231S,2013MNRAS.436..315F},
and to comprehend the interplay between its environment and the accretion
modes \citep{2013MNRAS.435..679F}.

Recently, galaxy clustering studies at intermediate and high redshift
\citep{2000MNRAS.317..782B,2003ApJ...588...50D,2006ApJ...644..671C,2009AA...505..463M,2014ApJ...793...17B,2014ApJ...784..128S}
have confirmed a strong correlation between galaxy luminosity and
clustering amplitude, previously found at lower redshifts (\citealp{1997ApJ...489...37G,2005ApJ...630....1Z,2010MNRAS.407...55L,2011ApJ...736...59Z}).
This suggests that most the luminous galaxies reside in more overdense
regions than less luminous ones. For quasar clustering, the picture
is less clear. Several authors have found a weak clustering dependency
on optical luminosity (e.g., \citealp{2005ApJ...619..697A,2005MNRAS.356..415C,2006MNRAS.371.1824P,2006ApJ...638..622M,2008MNRAS.383..565D,2011MNRAS.416..650S,2013ApJ...778...98S,2015MNRAS.453.2779E}).
These clustering results are in disagreement with the biased halo
clustering idea, in which more luminous quasars reside in the most
massive haloes, and therefore should have larger correlation lengths.
A weak dependency on the luminosity could imply that host halo mass
and quasar luminosity are not tightly correlated, and both luminous
and faint quasars reside in a broad range of host DMH masses. However,
these conclusions can be affected because the quasar samples are flux-limited,
and therefore often have small dynamical range in luminosity. In addition,
the intrinsic scatter for the different observables, such as the luminosity,
emission line width, and stars velocity dispersion, leads to uncertainties
in derivables such as halo, galaxy, and BH masses, which in turn could
mask any potential correlation between the observables and derivables.
For instance, \citet{2011ApJ...736..161C} assigned aleatory quasar
velocity widths to different objects and re-determined their BH masses.
They found that the differences between the randomized and original
BH masses are marginal. This implies that the low dispersion in broad-line
velocity widths provides little additional information to virial BH
mass estimations.

\citet{2009ApJ...697.1656S} divided their SDSS sample into bins corresponding
to different quasar properties: optical luminosity, virial BH mass,
quasar color, and radio-loudness. They found that the clustering strength
depends weakly on the optical luminosity and virial BH masses, with
the $10\%$ most luminous and massive quasars being more clustered
than the rest of the sample. Additionally, their radio-loud sample
shows a larger clustering amplitude than their radio-quiet sources.
Previous observations at low and intermediate redshift of the environments
of radio galaxies and radio-loud AGNs suggest that these reside in
denser regions compared with control fields (e.g., \citealp{2006ApJ...650L..29M,2013ApJ...769...79W}).
At $z\gtrsim1.5$, Mpc-sized dense regions have not yet virialized
within a single cluster-sized DMH and are consider to be the progenitors
of present day galaxy clusters \citep{2004ASSL..301..141K,2008A_ARv..15...67M}.
These results suggest that there is a relationship between radio-loud
AGNs and the environment in which these sources reside (see \citealt{2008A_ARv..15...67M}
for a review). 

Although the first known quasars were discovered as radio sources,
only a fraction of $\sim10\%$ are radio-loud \citep{1965ApJ...141.1560S}.
Radio-loud quasars (RLQs) and Radio-quiet quasars (RQQs) share similar
properties over a wide wavelength range of the electromagnetic spectrum,
from $100\:\mu m$ to the X-ray bands. The main difference between
both categories is the presence of powerful jets in RLQs (e.g. \citealt{1994AJ....108..766B,2008MNRAS.390..595M}).
However, there is evidence that RQQs have weak radio jets \citep{2005ApJ...621..123U,2006A_A...455..161L}.
How these jets form is still a matter of debate and their physics
is not yet completely understood. Several factors such as accretion
rate \citep{2010ApJ...723.1119L,2011MNRAS.411.1909F}, BH spin \citep{1977MNRAS.179..433B,2007ApJ...658..815S,2011MNRAS.411.1909F,2013AA...557L...7V},
BH mass \citep{2000ApJ...543L.111L,2003MNRAS.340.1095D,2011MNRAS.416..917C},
and quasar environment \citep{2001AJ....122.2833F,2013MNRAS.436..997R},
but most probably a combination of them, may be responsible for the
conversion of accreted material into well-collimated jets. This division
into RLQs and RQQs still remains a point of discussion. Some authors
advocate the idea that radio-loudness ($R$, radio-to-optical flux
ratio) distribution for optical-selected quasars is bimodal \citep{1989AJ.....98.1195K,1990MNRAS.244..207M,2002AJ....124.2364I,2007ApJ...656..680J},
while others have confirmed a very broad range for the radio-loudness
parameter, questioning its bimodality nature \citep{2003MNRAS.341..993C,2011ApJ...743..104S,2013ApJ...764...43S}. 

An important question in the study of the bimodality for the quasar
population is which physics sets the characteristic mass scale of
quasar host halos and the BHs that power them. Specifically, studying
the threshold for BH mass associated with the onset of significant
radio activity is crucial for addressing basic questions about the
physical process involved. According to the spectral analysis of homogeneous
quasar samples, RLQs are associated to massive BHs with $M_{\textrm{BH}}\gtrsim10^{9}$,
while RQQs are linked to BHs with $M_{\textrm{BH}}\lesssim10^{8}$
\citep{2000ApJ...543L.111L,2002MNRAS.336L..38J,2006MNRAS.365..101M}.
Other studies found that there is no such upper cutoff in the masses
for RQQs and they stretch across the full range of BH masses \citep{2002ApJ...576...81O,2002ApJ...581L...5W,2004MNRAS.353L..45M}. 

An alternative way to indirectly infer BH masses for radio-selected
samples is to use spatial clustering measurements. Most previous clustering
analyses for radio selected sources have found they are strongly clustered
with correlation lengths $r_{0}\sim11\,h^{-1}\,\textrm{Mpc}$ \citep{1991MNRAS.253..307P,1998MNRAS.300..257M,2003AA...405...53O}.
\citet{2004MNRAS.350.1485M} studied the clustering properties for
a sample of radio galaxies drawn from the Faint Images of the Radio
Sky at 20 cm (FIRST, \citealp{1995ApJ...450..559B}) and 2dF Galaxy
Redshift surveys (2dFGRS, \citealp{2001MNRAS.328.1039C}) and found
that they reside in typical DMH mass of $M_{\textrm{DMH}}\sim10^{13.4}\:\textrm{M}_{\odot}$,
with a BH mass of $\sim10^{9}\:\textrm{M}_{\odot}$, a value consistent
with BH mass estimations using composite spectra. A comparable limit
for the BH mass was found by \citet{2005MNRAS.362...25B} analyzing
a SDSS radio-AGN sample at low-z. Clustering measurements of the two-point
correlation function for RLQs (e.g. \citealp{2005MNRAS.356..415C,2009ApJ...697.1656S})
obtained $r_{0}$ values consistent with those of radio galaxies.
On the other hand, \citet{2010MNRAS.407.1078D} found that RLQs are
less clustered than radio galaxies, however, their sample was relative
smaller.

Clustering statistics offer an efficient way to explore the connections
between AGN types, including radio, X-ray, and infrared selected AGNs
\citep{2009ApJ...696..891H}; obscured and unobscured quasars \citep{2011ApJ...731..117H,2014ApJ...796....4A,2015arXiv151104469D};
radio galaxies \citep{2002MNRAS.333..100M,2008MNRAS.391.1674W,2011MNRAS.418.2251F};
blazars \citep{2014ApJ...797...96A}; and AGNs and galaxy populations:
Seyferts and normal galaxies; and optical quasars and submillimeter
galaxies \citep{2012MNRAS.421..284H}. These findings open up the
possibility to explain the validity and simplicity of unification
schemes (e.g. \citealp{1993ARAA..31..473A,1995PASP..107..803U}) for
radio AGNs with clustering.

The purpose of the present study is to measure the quasar clustering
signal, study its dependency on radio-loudness and BH virial mass,
and derive the typical masses for the host haloes and the SMBHs that
power these quasars. We use a sample of approximately $48000$ uniformly
selected spectroscopic quasars drawn from the SDSS DR7 (\citealt{2011ApJS..194...45S})
at $0.3\leq z\leq2.2$. In Section \ref{sec:Section2}, we present
our sample obtained from the joint use of the SDSS DR7 and FIRST surveys.
The methods used for the clustering measurement are introduced in
Section \ref{sec:Section3}. We discuss our results for the measurement
of the two-point correlation function for both RLQs and RQQs in Section
\ref{sec:Section4}. In addition, we compare our findings with previous
results from the literature. Finally, in Section \ref{sec:Section8},
we summarize our conclusions. Throughout this paper, we adopt a lambda
cold dark matter cosmological model with the matter density $\Omega_{m}=0.30$,
the cosmological constant $\Omega_{\Lambda}=0.70$, the Hubble constant
$H_{0}=70\,\textrm{km}\,\textrm{s}^{-1}\,\textrm{Mpc}^{-1},$ and
the rms mass fluctuation amplitude in spheres of size $8\,h^{-1}$
Mpc $\sigma_{8}=0.84$.

\section{Data \label{sec:Section2}}

\subsection{Sloan Digital Sky Survey}

The SDSS I/II was a photometric and spectroscopic survey of approximately
one-fourth of the sky using a dedicated wide-field $2.5\:\textrm{m}$
telescope \citep{1998AJ....116.3040G}. The resulting imaging provides
photometric observations in five bands: \textit{u, g, r, i,} and \textit{z}
\citep{1996AJ....111.1748F}. The selection for spectroscopic follow-up
for the quasars at low redshift ($z\leq3$) is done in the \textit{ugri
}color space with a limiting magnitude of $i\leq19.1$ \citep{2002AJ....123.2945R}.
At high-redshift ($z\geq3$), the selection is performed in \textit{griz
}color space with $i<20.2$. The quasar candidates are assigned to
$3^{\circ}$ diameter spectroscopic plates by a tiling algorithm \citep{2003AJ....125.2276B}
and observed with double spectrographs with a resolution of $\lambda/\Delta\lambda\sim2000$.
Each plate hosts $640$ fibers and two fibers cannot be closer than
$55^{\prime\prime}$, which corresponds to a projected distance of
$0.6-1.5\,h^{-1}\:\textrm{Mpc}$ for $0.3<z<2.3$. This restriction
is called fiber collisions, and causes a deficit of quasar pairs with
projected separations $\leq2\;\textrm{Mpc}$. We did not attempt to
compensate for pair losses due to fiber collisions, therefore we only
model our results for projected distances $\geq2\:\textrm{Mpc}$.

\noindent We exploit the \citet{2011ApJS..194...45S} value-added
catalog that is based on the main SDSS DR7 quasar parent sample \citet{2010AJ....139.2360S}.
We select a flux limited $i=19.1$ sample of $48338$ quasars with
$0.3\leq z\leq2.3$ from the \citet{2011ApJS..194...45S} catalog
with the flag\textsc{ uniform\_target$=1$. }This sample includes
both RLQs and RQQs selected uniformly by the quasar target selection
algorithm presented in \citet{2002AJ....123.2945R}. For quasar clustering
studies, it is critical to use statistical samples that have been
constructed using only one target selection algorithm. Therefore,
this sample excludes SDSS objects with non-fatal photometric errors
and are selected for spectroscopic follow-up based only on their radio
detection in the FIRST survey (see \citealt{2002AJ....123.2945R}
for more details). The combination of quasars selected employing different
target selections could lead to the appearance of potential systematics
in the resulting sample. This includes higher clustering strength
at large scales \citep{2009ApJ...697.1634R}. Previous studies using
uniform samples have shown that these are very stable and insensitive
to systematic effects such as dust reddening, and bad photometry \citep{2009ApJ...697.1634R,2009ApJ...697.1656S,2013ApJ...778...98S}. 

\subsection{FIRST survey}

\begin{figure}[h]
\centering{}\centering\includegraphics[bb=0bp 0bp 424bp 360bp,scale=0.6]{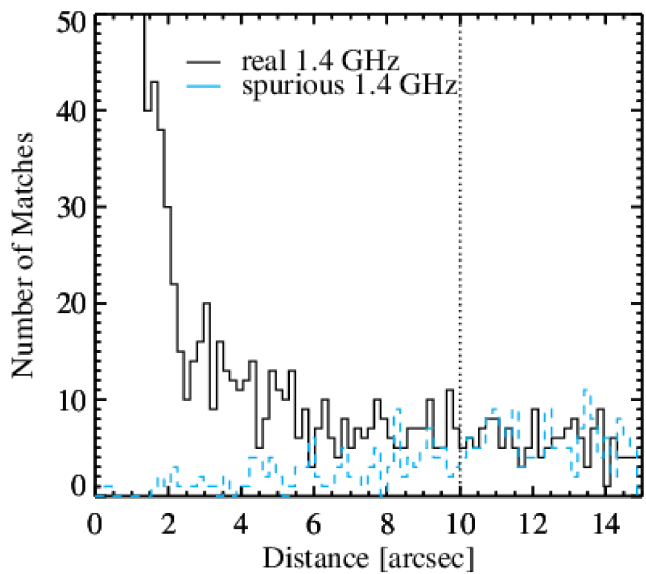}\caption{\label{fig:Matching-histogram}The solid histogram shows the distance
distribution for SDSS quasar counterparts to $\textrm{S}_{1.4\:\textrm{GHz}}\geq1.0\:\textrm{mJy}$
FIRST radio sources. Cyan dashed histogram indicates the distribution
for spurious associations, which are obtained by vertically shifting
the quasar positions by $1^{\prime}$.}
\end{figure}

\noindent 
\begin{figure*}[t]
\centering{}\centering\includegraphics[scale=0.5]{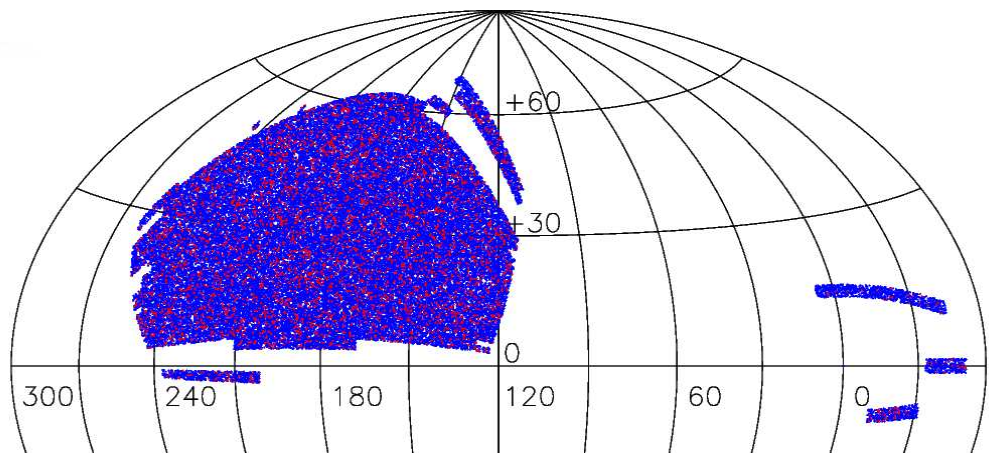}\caption{\label{fig:Geometry}Aitoff projection for the sky coverage of the
SDSS DR7 uniform quasar sample from \citet{2011ApJS..194...45S}.
RQQs are denoted by blue points, while the RLQs are represented by
red points. See Section \ref{sec:Section2} for a description of the
methodology employed in the selection for the RLQs.}
\end{figure*}

\noindent 
\begin{figure}
\centering{}\centering\includegraphics[scale=0.34,clip=true]{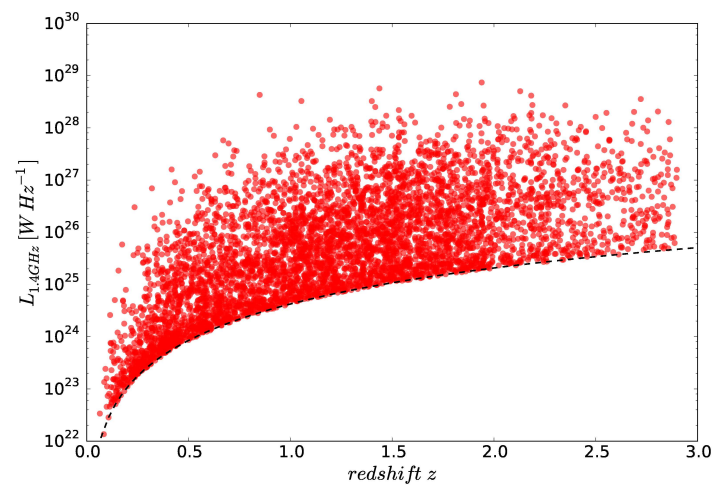}\caption{\label{fig:radioinfo} The $1.4$ GHz restframe radio luminosity for
the RLQs (red) detected in the FIRST radio survey. We assume a radio
spectral index of $0.70$, and a flux limit of $1.0\:\textrm{mJy}$.
The dashed lines show the luminosity limit for the FIRST survey flux
limit. }
\end{figure}

The FIRST survey \citep{1995ApJ...450..559B} is a radio survey at
1.4 GHz that aims to map $10000$ square degrees of the North and
South Galactic Caps using the NRAO Very Large Array. The FIRST radio
observations are done using the B-array configuration providing an
angular resolution of $\sim5^{\prime\prime}$ with positional accuracy
better than $1^{\prime\prime}$ at a limiting radio flux density of
$1\:\textrm{mJy}$ ($5\,\sigma$) for point sources. FIRST was designed
to have an overlap with the SDSS survey, and yields a $40\%$ identification
rate for optical counterparts at the $m_{V}\sim23$ (SDSS limiting
magnitude).

\subsection{Cross-matching of the SDSS and FIRST catalogs}

The quasar catalog provided by \citet{2010AJ....139.2360S} is matched
to the FIRST catalog taking sources with position differences less
than $2^{\prime\prime}$. However, this short distance prevents the
identification of quasars with diffuse or complex radio emission.
Therefore, to account for RLQs possibly missed by the original matching,
we cross-matched the SDSS and FIRST catalogs with larger angular distances.
To choose the upper limit for a new matching radius, we vertically
shifted the quasar positions by $1^{\prime}$ and proceeded to match
again with the FIRST catalog. Shown by a solid line in Fig. \ref{fig:Matching-histogram}
we reproduce the distribution of angular distances between SDSS objects
and their nearest FIRST counterpart, and by a dashed line the we show
distribution of spurious matches. The distribution of real matches
presents a peak and a declining tail that flattens with increasing
distance. Both distributions are at the same level at $\sim10^{\prime\prime}$.
This radius will be used as the maximum angular separation for matching
the SDSS and FIRST surveys. This value is a good compromise between
the maximum number of real identifications and keeping the spurious
associations to a minimum. The total number of newly identified radio
quasars with angular offsets between $2^{\prime\prime}$ and $10^{\prime\prime}$
is 409. 

\noindent Some statistical matching methods, such as the likelihood
ratio (LR), have been proposed to robustly cross-match radio and optical
surveys (e.g., \citealt{1992MNRAS.259..413S}). \citet{2004ApJS..155....1S}
showed that when the positional uncertainties for both radio and optical
catalogues are small, the LR technique and positional coincidence
yield very similar results. This is the case for both catalogs used
in this work, which have accurate astrometry ($\sim0.1^{\prime\prime}$
for SDSS, $\sim1^{\prime\prime}$ for FIRST). The contamination rate
by random coincidences \citep{2007AN....328..577E,2014MNRAS.440.1527L}
is:

\noindent \begin{center}
\begin{equation}
P_{\textrm{C}}=\pi\,r_{\textrm{s}}^{2}\rho,\label{eq:}
\end{equation}
\par\end{center}

\noindent where $r_{\textrm{s}}$ is the matching radius, and $\rho\simeq5.6\,\textrm{deg}^{-2}$
is the quasar surface density. For $r_{\textrm{s}}=2^{\prime\prime}$,
the expected number of contaminants in the RLQs sample is $2$, while
for $r_{\textrm{s}}=10^{\prime\prime}$ this rate increases to $61$.
This small contamination fraction ($<2\%$ from the total radio sample)
is unlikely to affect our clustering measurements. 

\noindent The sensibility for the FIRST survey is not uniform across
the sky, with fluctuations due to different reasons, such as hardware
updates, observing strategies, target declination, and increasing
noise in the neighborhood of bright sources \citep{1995ApJ...450..559B}.
Despite all these potential limitations, the detection limit for most
of the targeted sky is a peak flux density of $1\:\textrm{mJy}$ $\left(5\sigma\right)$,
with only an equatorial strip having a slightly deeper detection threshold
due to the combination of two observing epochs. We refer the interested
reader to \citet{2015ApJ...801...26H}, where the impact of all the
above mentioned aspects is discussed extensively. The flux limit of
$1\:\textrm{mJy}$ is considered only for peak flux density instead
of integrated flux density. Hence a source with peak fluxes individually
smaller than the detection threshold but with total flux greater than
this value could not appear in our radio sample. In particular, lobe-dominated
quasars (see \citealt{1974MNRAS.167P..31F}; hereafter FR2) with peak
fluxes less than the flux limit suffer from a systematic incompleteness
in comparison to core-dominated quasars (FRI). We investigate how
not taking into account FIRST resolution effects could possibly affect
our RLQ clustering measurements. We estimate the weights for RLQs
with fluxes less than $5\:\textrm{mJy}$ using the completeness curve
from \citet{2007AJ....134.1150J} , which takes into account the source
morphology and rms values in the FIRST survey for SDSS quasars. We
find that including a weighting scheme does not affect the clustering
signal for RLQs.

\begin{figure}[h]
\centering{}\centering\includegraphics[bb=8bp 10bp 560bp 432bp,clip,scale=0.47]{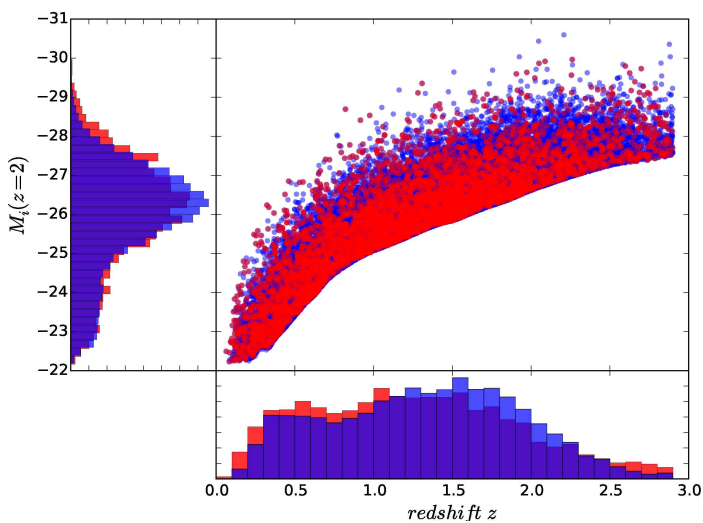}\caption{\label{fig:MI_Z2_redshift} The distribution of RLQs (red) and RQQs
(blue) in the optical-luminosity space. The absolute magnitude in
the i-band at z = 2 $M_{i}(z=2)$ is calculated using the K-correction
from \citet{2006AJ....131.2766R}. The left and bottom panels show
the $M_{i}(z=2)$ and redshift histograms. The normalized redshift
and optical-luminosity distributions are displayed in the left and
bottom panels. The normalized distributions for both samples show
a good degree of similarity, allowing a direct comparison of their
clustering measurements.}
\end{figure}

We define a quasar to be radio-loud if it has a detection in the FIRST
with a flux above $1\:\textrm{mJy}$, and radio-quiet if it is undetected
in the radio survey. To minimize incompleteness due to the FIRST flux
limit while retaining the maximum numbers of quasars for clustering
measurements, we consider two radio-luminosity cuts: $L_{1.4\,\textrm{GHz}}>4\times10^{24}\,W\,Hz^{-1}$
for $0.3<z<1.0$; and $L_{1.4\,\textrm{GHz}}>1\times10^{25}\,W\,Hz^{-1}$
for $1.0<z<2.3$. Our parent sample then comprises a total of $45441$
RQQs and $3493$ RLQs with $0.3<z<2.3$, which corresponds to a radio-loud/-quiet
source fraction of $\sim7.2\%$. This ratio is in agreement with previous
studied quasar samples (e.g., \citealp{2007ApJ...656..680J,2011AJ....142....3H}).
This choice for the redshift range avoids the poor completeness at
high-z due to color confusion with stars in the \textit{ugri }color
cube. The sky coverage of our final quasar sample of $6248\;\textrm{deg}^{2}$
is shown in Fig. \ref{fig:Geometry}. We calculate the radio-luminosity
adopting a mean radio spectral index of $\alpha_{\textrm{rad}}=0.7$
(where $S_{\nu}\varpropto\nu^{-\alpha}$ ) and applying the usual
k-correction for the luminosity estimation. Fig. \ref{fig:radioinfo}
shows the radio-luminosity for our quasar sample. The quasar distribution
in the optical-luminosity redshift plane is displayed in Fig. \ref{fig:MI_Z2_redshift}.
The normalized redshift and optical-luminosity distributions for both
samples show a good degree of similarity, this allows a direct comparison
of their clustering measurements. We confirm this by applying two
Kolmogorov-Smirnov (K-S) tests, which indicate a probability for the
redshift and luminosity redshift distributions of 95\% and 97\%, respectively,
that both samples (RLQs and RQQs) are drawn from the same parent distribution. 

\subsection{Final quasar sample \label{sec:Section2.4-Quasarsample}}

\begin{figure}
\centering{}\centering\includegraphics[scale=0.34,clip=true]{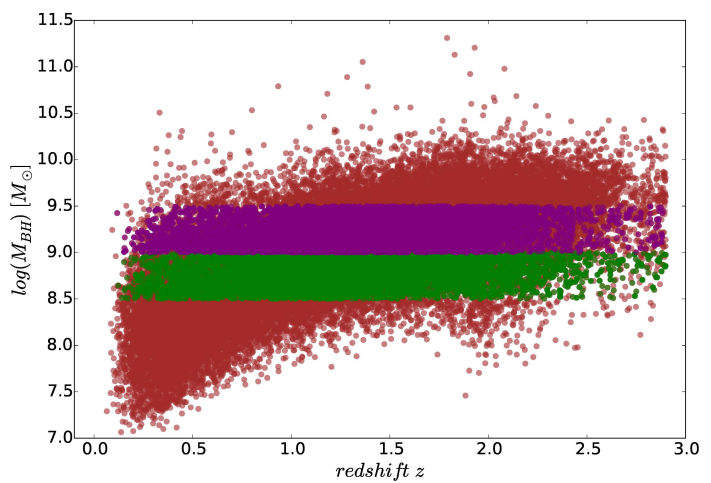}\caption{\label{fig:OptLuminosities-BHMasses} The quasar distribution in the
virial BH mass plane. The quasars selected to match in optical luminosity
with masses $8.5\leq\log\left(M_{\textrm{BH}}\right)\leq9.0$ are
indicated with green color, and the objects with $9.0\leq\log\left(M_{\textrm{BH}}\right)\leq9.5$
are represented by purple points. The properties of the mass
samples are summarized in Table \ref{tab:quasar_best_fitting_values}.}
\end{figure}

The final spectroscopic quasar sample restricted to $0.3<z<2.3$ provides
an excellent dataset for probing the clustering dependence based on
physical properties such as radio-loudness or BH virial mass. It is
possible to explore how clustering depends on these properties to
some degree across different redshift intervals. Previous quasar clustering
studies (e.g., \citealt{2005MNRAS.356..415C,2009ApJ...697.1634R,2009ApJ...697.1656S})
were limited by their sample size ($\lesssim30000$ quasars) and studied
the correlation function for RLQs in only one redshift bin corresponding
to the entire redshift range of the sample. We take advantage of the
higher quasar numbers of our sample and divide each redshift bin into
smaller bins using radio-loudness and the virial BH masses as indicators,
and still obtain a good S/N for the correlation function of the samples
in our analysis. The $M_{\textrm{BH}}-z$ space is not uniformly populated.
We limit our analysis to two mass samples that are separated according
to their BH mass: $8.5\leq\log\left(M_{\textrm{BH}}\right)\leq9.0$
and $9.0\leq\log\left(M_{\textrm{BH}}\right)\leq9.5$. The redshift
distributions for these two mass bins are very different, with more
massive BHs peaking at $z\sim2$, while less massive at $z\sim0.5$
(see Fig. \ref{fig:OptLuminosities-BHMasses}). This hampers a direct
comparison between their clustering measurements. Thus, we create
control samples by randomly selecting quasars from the initial BH
mass samples that are matched by their optical luminosity distribution.
We verify that the resulting samples can be compared by applying a
K-S test to the new redshift distributions. This indicates a probability
of 97\% that the mass samples are drawn from the same parent distribution.
The properties for all the quasar samples are presented in Table \ref{tab:quasar_best_fitting_values}. 

\begin{singlespace}

\begin{table}
\caption{Main properties of our quasar samples. The bar denotes the median
values.\label{tab:quasar_sample}}

\centering{}%
\begin{tabular}{cccc}
 &  &  & \tabularnewline
\hline 
{\tiny{}Sample } & {\tiny{}$\bar{M}_{\textrm{BH}}$ } & {\tiny{}$\bar{L}_{\textrm{Bol}}$ } & {\tiny{}$\bar{L}_{1.4\,GHz}$ }\tabularnewline
 & {\tiny{} $[\log\left(M_{\odot}\right)]$ } & {\tiny{}$[10^{46}\,\textrm{erg}\,\textrm{s}^{-1}]$ } & {\tiny{}$[10^{26}\,\textrm{W}\,\textrm{Hz}^{-1}]$ }\tabularnewline
\hline 
{\tiny{}$0.3\leq z\leq2.3$ } &  &  & \tabularnewline
{\tiny{}All } & {\tiny{}$9.21$} & {\tiny{}$4.72$} & {\tiny{}-}\tabularnewline
{\tiny{}RQQs} & {\tiny{}$9.19$ } & {\tiny{}$3.57$} & {\tiny{}-}\tabularnewline
{\tiny{}RLQs} & {\tiny{}$9.36$} & {\tiny{}$5.69$} & {\tiny{}$8.32$}\tabularnewline
{\tiny{}$9.0\leq\log(M_{\textrm{BH}})\leq9.5$} & {\tiny{}$9.23$} & {\tiny{}$1.48$} & {\tiny{}-}\tabularnewline
{\tiny{}$8.5\leq\log(M_{\textrm{BH}})\leq9.0$} & {\tiny{}$8.82$} & {\tiny{}$2.14$} & {\tiny{}-}\tabularnewline
\hline 
{\tiny{}$0.3\leq z\leq1.0$ } &  &  & \tabularnewline
{\tiny{}RQQs} & {\tiny{}$8.80$} & {\tiny{}$0.90$} & {\tiny{}-}\tabularnewline
{\tiny{}RLQs} & {\tiny{}$9.35$} & {\tiny{}$6.43$} & {\tiny{}$2.54$}\tabularnewline
{\tiny{}$9.0\leq\log(M_{\textrm{BH}})\leq9.5$} & {\tiny{}$9.20$} & {\tiny{}$0.79$} & {\tiny{}-}\tabularnewline
{\tiny{}$8.5\leq\log(M_{\textrm{BH}})\leq9.0$} & {\tiny{}$8.77$} & {\tiny{}$0.85$} & {\tiny{}-}\tabularnewline
\hline 
{\tiny{}$1.0\leq z\leq2.3$ } &  &  & \tabularnewline
{\tiny{}RQQs } & {\tiny{}$9.15$} & {\tiny{}$4.70$} & {\tiny{}-}\tabularnewline
{\tiny{}RLQs} & {\tiny{}$9.07$} & {\tiny{}$5.39$} & {\tiny{}$10.6$}\tabularnewline
{\tiny{}$9.0\leq\log(M_{\textrm{BH}})\leq9.5$} & {\tiny{}$9.23$} & {\tiny{}$2.57$} & {\tiny{}-}\tabularnewline
{\tiny{}$8.5\leq\log(M_{\textrm{BH}})\leq9.0$} & {\tiny{}$8.84$} & {\tiny{}$2.69$} & {\tiny{}-}\tabularnewline
\hline 
 &  &  & \tabularnewline
\end{tabular}
\end{table}

\end{singlespace}

\noindent 

\section{Clustering of quasars\label{sec:Section3}}

\subsection{Two-point correlation functions}

\begin{figure}[h]
\raggedright{}\centering\includegraphics[scale=0.42,clip=true]{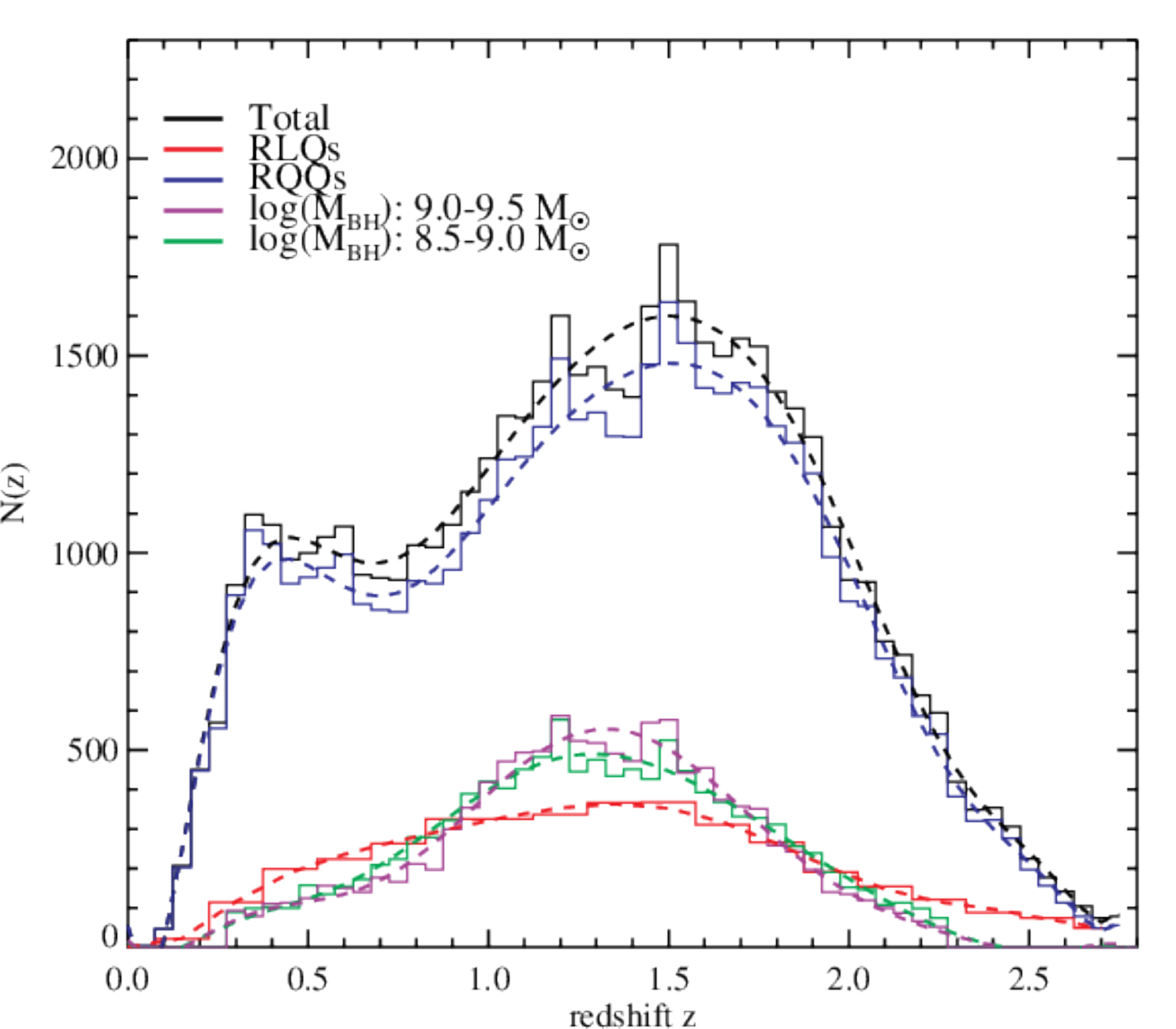}\caption{\label{fig:histograms}Redshift distributions for the total quasar
sample (black), RQQs (blue), RLQs (red), quasars with $8.5\leq\log\left(M_{\textrm{BH}}\right)\leq9.0$
(green) and $9.0\leq\log\left(M_{\textrm{BH}}\right)\leq9.5$ (purple).
The mass samples are matched in optical luminosity at each redshift
interval (see Section \ref{sec:Section4-1} for more details). The
solid lines are fitted polynomials used to generate the random quasar
catalogs used in the correlation function estimations. }
\end{figure}

The two-point correlation function (TPCF) $\xi\left(r\right)$ describes
the excess probability of finding a quasar at a redshift distance
$r$ from a quasar selected randomly over a random distribution. To
contraint this function, we create random catalogs with the same angular
geometry and the same redshift distribution as the data with at least
$70$ times the number of quasars in the data sets to minimize the
impact of Poisson noise. The redshift distributions corresponding
to the different quasar samples are shown by the solid lines in Fig.
\ref{fig:histograms}.

\noindent 

\noindent The TPCF is estimated using the minimum variance estimator
suggested by \citet{1993ApJ...412...64L} 

\noindent \begin{center}
\begin{equation}
\xi_{\textrm{LS}}=\frac{DD-2\,DR+RR}{RR},\label{eq:LS-estimator}
\end{equation}
\par\end{center}

\noindent where $DD$ is the number of distinct data pairs, $RR$
is the number of different random pairs, and $DR$ is the number of
cross-pairs between the real and random catalogs within the same bin.
All pair counts are normalized by $n_{QSO}$ and $n_{R}$, respectively,
the mean number densities in the quasar and random catalogs. We verify
our estimates using the Hamilton estimator \citep{1993ApJ...417...19H},
and find a good agreement of the results for both estimators within
the error bars, although the LS estimator is preferred because it
is less sensitive to edge effects.

\noindent In reality, observed TPCFs are distorted both at large and
small scales. On smaller scales, quasars have peculiar non-linear
velocities that cause an elongation along the line of sight, which
is referred as the Finger of God effect \citep{1972MNRAS.156P...1J}.
At larger scales, the coherent motion of quasars that are infalling
onto still-collapsing structures produces a flattening of the clustering
pattern to the observer. This distortion is called the Kaiser effect
\citep{1987MNRAS.227....1K}. 

\noindent Because of the existing bias mentioned earlier in redshift-space,
a different approach is used to minimize the distortion effects in
the clustering signal \citep{1983ApJ...267..465D}. Following \citet{1994MNRAS.266...50F},
we use the separation vector, $\mathbf{s}=\mathbf{s}_{1}-\mathbf{s}_{2}$,
and the line of sight vector, $\mathbf{l}=\mathbf{s}_{1}+\mathbf{s_{2}}$;
where $\mathbf{s}_{1}$ and $\mathbf{s}_{2}$ are the redshift-space
position vectors. From these, it is possible to define the parallel
and perpendicular distances for the pairs as:

\begin{eqnarray}
\pi=\frac{\left|\mathbf{s\cdot l}\right|}{\left|\mathbf{l}\right|}, &  & r_{p}=\sqrt{\mathbf{s\cdot s}-\pi^{2}}.
\end{eqnarray}

\noindent Now, we can compute the correlation function $\xi\left(r_{p},\pi\right)$
in a two-dimensional grid using the LS estimator, as in eq. \eqref{eq:LS-estimator}.
Because the redshift distortions only affect the distances in the
$\mathbf{\pi-\textrm{direction}}$, we integrate along this component
and project it on the $r_{p}-\textrm{axis}$ to obtain the projected
correlation function

\noindent \begin{center}
\begin{equation}
\frac{w_{p}\left(r_{p}\right)}{r_{p}}=\frac{2}{r_{p}}\,\int_{0}^{\infty}\,\xi\left(r_{p},\pi\right)\,\textrm{d}\pi,\label{eq:projected-tpcf}
\end{equation}
\par\end{center}

\noindent which is independent of redshift-space distortions, as it
measures the clustering signal as a function of the quasar separation
in the perpendicular direction to the line of sight.

\begin{figure}[h]
\centering{}\centering\includegraphics[scale=0.46]{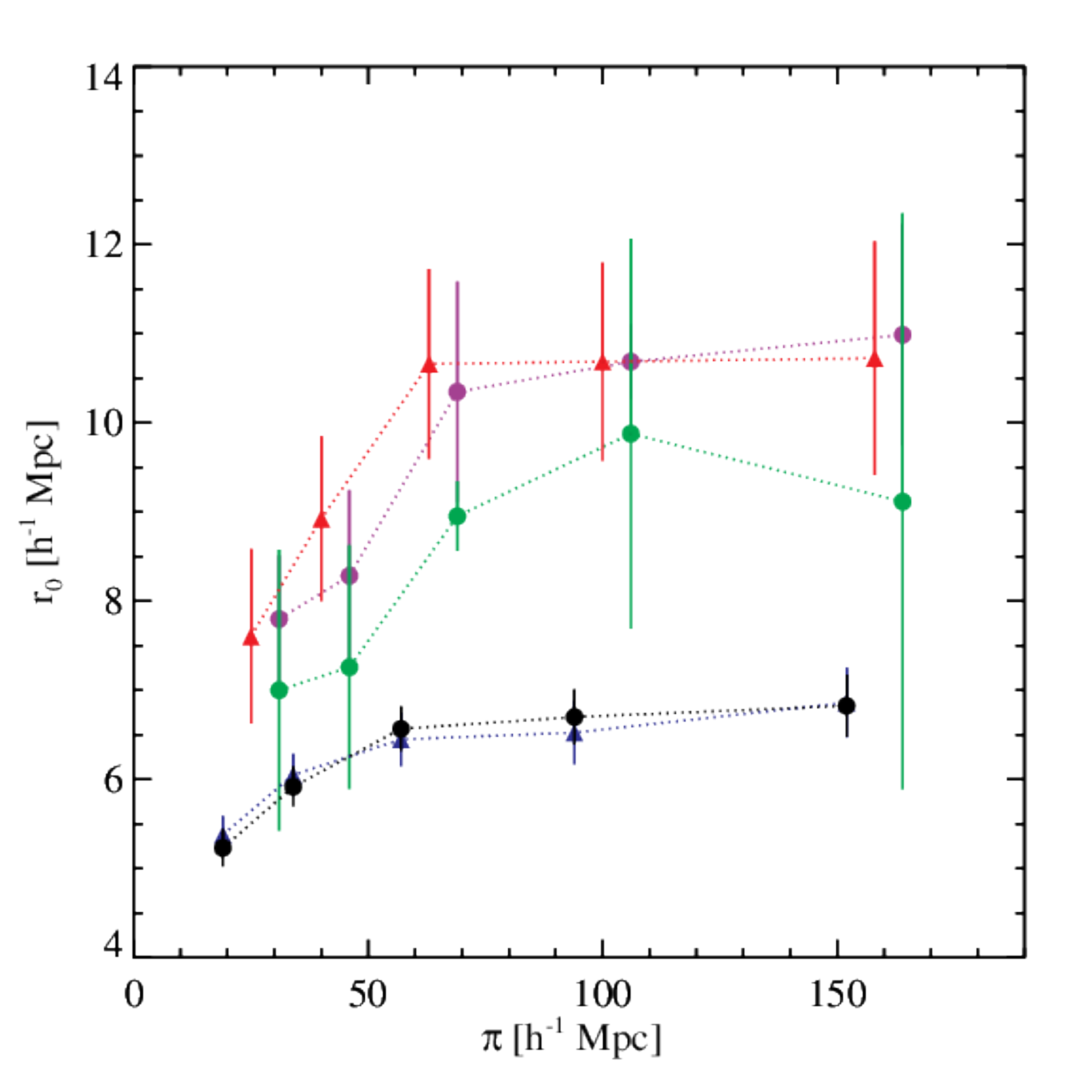}\caption{\label{fig:r0_vs_pimax} Real-space correlation length $r_{0}$
vs the parallel direction to the line of sight $\pi$ for the full
quasar sample (black circles), $9.0\leq\log\left(M_{\textrm{BH}}\right)\leq9.5$
sample (purple circles), $8.5\leq\log\left(M_{\textrm{BH}}\right)\leq9.0$
sample (green circles), RQQs (blue triangles), and RLQs (red triangles).
For clarity, the mass samples have been shifted by $\pi=6\,h^{-1}\,\textrm{Mpc}$,
and the full and RQQs samples by $\pi=6\,h^{-1}\,\textrm{Mpc}$. }
\end{figure}

\noindent In practice, it is not feasible to integrate eq. \eqref{eq:projected-tpcf}
to infinity, thus an upper limit $\pi_{max}$ to the integral shall
be chosen to be a good compromise between the impact of noise and
a reliable calculation of the measured signal. We try several $\pi$
upper limits by fitting $w_{p}$ to a power-law of the form \citep{1983ApJ...267..465D} 
\noindent \begin{center}
\begin{equation}
w_{p}\left(r_{p}\right)=r_{p}\left(\frac{r_{0}}{r_{p}}\right)^{\gamma}\left[\frac{\Gamma\left(\frac{1}{2}\right)\Gamma\left(\frac{\gamma-1}{2}\right)}{\Gamma\left(\frac{\gamma}{2}\right)}\right],\label{eq:power-law}
\end{equation}
,
\par\end{center}

\noindent where $r_{0}$ is the real-space correlation length, and
$\gamma$ the power-law slope. We use the range $2.0\leq r_{p}\leq130\,h^{-1}\:\textrm{Mpc}$
to determine the scale at which the clustering signal is stable (Fig.
\ref{fig:r0_vs_pimax}). We find that above $\pi=63.1\,h^{-1}\,\textrm{Mpc}^{-1},$
the fluctuations in the correlation length are within uncertainties
and have poorer S/N. Thus, we take this value as our upper integration
limit $\pi_{max}$, which is within the range $40-70\,h^{-1}\,\textrm{Mpc}^{-1}$
of previous quasar clustering studies (e.g. \citealt{2004MNRAS.355.1010P,2009ApJ...697.1634R}). 

\subsection{Error estimation \label{sec:Error-estimation}}

We calculate the errors from the data itself by using the delete-one
jackknife method \citep{2009MNRAS.396...19N}. We divide the survey
into $N_{sub}$ different sub-samples, and \emph{delete} one sample
at a time to compute the correlation function for $N_{sub}-1$ sub-samples.
This process is repeated $N_{sub}$ times to obtain the correlation
function for bin $i$ in the jackknife sub-sample $k$, denoted by
$\xi_{i}^{k}$. We can write the jackknife covariance matrix (e.g.
\citealp{2002ApJ...579...48S,2009MNRAS.396...19N}) as

\noindent \begin{center}
\begin{equation}
C_{ij}=\frac{N_{sub}-1}{N_{sub}}\,\sum_{k=1}^{N_{sub}}\left(\xi_{i}^{k}-\xi_{i}\right)\left(\xi_{j}^{k}-\xi_{j}\right),
\end{equation}
\par\end{center}

\noindent with $\xi_{i}$ the correlation function for all data at
each bin $i$. We employ a total of $N_{sub}=24$ sub-samples for
our error estimations. Each sub-sample is chosen to be an independent
cosmological volume with approximately the same number of quasars.
The off-diagonal elements in the covariance matrix are small at large
scales and could potentially insert some noise into the inverse matrix
\citep{2009ApJ...697.1634R,2009ApJ...697.1656S}. Therefore, we employ
only diagonal elements for the $\chi^{2}$ fitting.

\subsection{Bias, dark matter halo and black hole mass estimations \label{subsec:mass_estimations}}

According to the linear theory of structure formation, the bias parameter
$b$ relates the clustering amplitude of large-scale structure tracers
and the underlying dark matter distribution. The quasar bias parameter
can be defined as 

\noindent \begin{center}
\begin{equation}
b^{2}=w_{\textrm{QSO}}\left(r_{p},z\right)/w_{\textrm{DM}}\left(r_{p},z\right),
\end{equation}
\par\end{center}

\noindent where $w_{\textrm{QSO}}$ and $w_{\textrm{DM}}$ are the
quasar and dark matter correlation functions \citep{1980lssu.book.....P},
respectively. We estimate the bias factor using the halo model approach,
in which $w_{\textrm{DM}}$ has two contributions: the 1-halo and
2-halo terms. The first term is related to quasar pairs from within
the same halo, and the second one is the contribution from quasars
pairs in different haloes. As the latter term dominates at large separations,
we can neglect the 1-halo term and write $w_{\textrm{DM}}$ as \citep{2002ApJ...568..455H}

\noindent \begin{center}
\begin{equation}
w_{\textrm{DM}}\left(r_{p},z\right)=w_{\textrm{DM}}^{2-h}\left(r_{p},z\right)=r_{p}\,\int_{r_{p}}^{\infty}\,\frac{r\,\xi_{\textrm{DM}}^{2-h}\left(r\right)}{\sqrt{r^{2}-r_{p}^{2}}}\,\textrm{d}r,\label{eq:wdm_model}
\end{equation}
\par\end{center}

\noindent with 

\noindent \begin{center}
\begin{equation}
\xi_{\textrm{DM}}^{2-h}\left(r\right)=\frac{1}{2\,\pi^{2}}\,\int\,P^{2-h}\left(k\right)\,k^{2}\,j_{0}\left(kr\right)\,\textrm{dk},
\end{equation}
\par\end{center}

\noindent where $k$ is the wavelength number, $h$ refers to the
halo term, $P^{2-h}\left(k\right)$ is the Fourier transform of the
linear power spectrum \citep{1992MNRAS.258P...1E} and $j_{0}\left(x\right)$
is the spherical Bessel function of the first kind. 

\noindent With the bias factor, it is possible to derive the typical
mass for the halo in which the quasars reside. We follow the procedure
described in previous AGN clustering studies (e.g., \citealt{2007ApJ...658...85M,2010ApJ...713..558K,2014ApJ...796....4A})
using the ellipsoidal gravitational collapse model of \citet{2001MNRAS.323....1S}
and the analytical approximations of \citet{2002MNRAS.331...98V}.

\section{Results \label{sec:Section4}}

\begin{singlespace}

\begin{table*}
\caption{Best-fitting correlation function model parameters for the quasar
samples. The range for the fits is $2.0\leq r\leq130\,h^{-1}\:\textrm{Mpc}$.
\label{tab:quasar_best_fitting_values}}

\centering{}%
\begin{tabular}{ccccccccc}
 &  &  &  &  &  &  &  & \tabularnewline
\hline 
{\tiny{}Sample } & {\tiny{}$\bar{z}$} & {\tiny{}$\textrm{N}_{\textrm{QSO}}$ } & {\tiny{}$r_{0}$} & {\tiny{}$\gamma$} & {\tiny{}$\chi^{2}$} & {\tiny{}$\textrm{DOF}$ } & {\tiny{}$b$} & {\tiny{}$\,M_{\textrm{DMH}}$}\tabularnewline
 &  &  & {\tiny{}{[}$h^{-1}\textrm{Mpc}${]} } &  &  &  &  & {\tiny{} {[}$h^{-1}M_{\odot}${]}}\tabularnewline
\hline 
{\tiny{}$0.3\leq z\leq2.3$ } &  &  &  &  &  &  &  & \tabularnewline
{\tiny{}All } & {\tiny{}$1.30$} & {\tiny{}$48338$} & {\tiny{}$6.81_{-0.30}^{+0.29}$} & {\tiny{}$2.10_{-0.05}^{+0.05}$} & {\tiny{}$20.17$} & {\tiny{}$7$} & {\tiny{}$2.00\pm0.08$} & {\tiny{}$2.33_{-0.38}^{+0.41}\times10^{12}$}\tabularnewline
{\tiny{}RQQs} & {\tiny{}$1.30$ } & {\tiny{}$45441$} & {\tiny{}$6.59_{-0.24}^{+0.33}$} & {\tiny{}$2.09_{-0.09}^{+0.10}$} & {\tiny{}$19.60$} & {\tiny{}$7$} & {\tiny{}$2.01\pm0.08$ } & {\tiny{}$2.38_{-0.38}^{+0.42}\times10^{12}$}\tabularnewline
{\tiny{}RLQs} & {\tiny{}$1.32$} & {\tiny{}$3493$} & {\tiny{}$10.95_{-1.58}^{+1.22}$} & {\tiny{}$2.29_{-0.34}^{+0.53}$} & {\tiny{}$1.06$} & {\tiny{}$7$} & {\tiny{}$3.14\pm0.34$ } & {\tiny{}$1.23_{-0.39}^{+0.47}\times10^{13}$}\tabularnewline
{\tiny{}$8.5\leq\log(M_{\textrm{BH}})\leq9.0$} & {\tiny{}$1.31$ } & {\tiny{}$11356$} & {\tiny{}$8.53_{-2.25}^{+1.57}$} & {\tiny{}$1.84_{-0.20}^{+0.21}$} & {\tiny{}$0.69$} & {\tiny{}$6$} & {\tiny{}$2.64\pm0.42$ } & {\tiny{}$6.57_{-0.31}^{+0.43}\times10^{12}$}\tabularnewline
{\tiny{}$9.0\leq\log(M_{\textrm{BH}})\leq9.5$} & {\tiny{}$1.31$ } & {\tiny{}$11356$} & {\tiny{}$10.45_{-0.98}^{+0.79}$} & {\tiny{}$2.36_{-0.17}^{+0.18}$} & {\tiny{}$1.99$} & {\tiny{}$6$} & {\tiny{}$2.99\pm0.43$ } & {\tiny{}$1.02_{-0.42}^{+0.55}\times10^{13}$}\tabularnewline
\hline 
{\tiny{}$0.3\leq z\leq1.0$ } &  &  &  &  &  &  &  & \tabularnewline
{\tiny{}RQQs} & {\tiny{}$0.65$} & {\tiny{}$13219$} & {\tiny{}$6.85_{-0.40}^{+0.45}$} & {\tiny{}$2.04_{-0.07}^{+0.08}$} & {\tiny{}$2.74$} & {\tiny{}$7$} & {\tiny{}$1.52\pm0.09$} & {\tiny{}$3.53_{-0.91}^{+1.07}\times10^{12}$}\tabularnewline
{\tiny{}RLQs} & {\tiny{}$0.71$ } & {\tiny{}$1019$} & {\tiny{}$18.39_{-2.01}^{+1.75}$} & {\tiny{}$2.40_{-0.16}^{+0.19}$} & {\tiny{}$1.95$} & {\tiny{}$4$} & {\tiny{}$4.63\pm0.58$ } & {\tiny{}$1.16_{-0.33}^{+0.37}\times10^{14}$}\tabularnewline
{\tiny{}$8.5\leq\log(M_{\textrm{BH}})\leq9.0$} & {\tiny{}$0.74$} & {\tiny{}$2604$} & {\tiny{}$10.90_{-2.48}^{+1.97}$} & {\tiny{}$1.54_{-0.14}^{+0.15}$} & {\tiny{}$0.54$} & {\tiny{}$6$} & {\tiny{}$2.83\pm0.45$} & {\tiny{}$2.89_{-1.23}^{+1.56}\times10^{13}$}\tabularnewline
{\tiny{}$9.0\leq\log(M_{\textrm{BH}})\leq9.5$} & {\tiny{}$0.74$ } & {\tiny{}$2604$} & {\tiny{}$15.26_{-2.09}^{+2.15}$} & {\tiny{}$2.29_{-0.36}^{+0.56}$} & {\tiny{}$1.11$} & {\tiny{}$6$} & {\tiny{}$3.56\pm1.02$ } & {\tiny{}$5.59_{-3.53}^{+5.10}\times10^{13}$}\tabularnewline
\hline 
{\tiny{}$1.0\leq z\leq2.3$ } &  &  &  &  &  &  &  & \tabularnewline
{\tiny{}RQQs } & {\tiny{}$1.58$ } & {\tiny{}$31102$} & {\tiny{}$6.61_{-0.70}^{+0.80}$} & {\tiny{}$2.13_{-0.09}^{+0.10}$} & {\tiny{}$14.70$} & {\tiny{}$6$} & {\tiny{}$2.21\pm0.10$ } & {\tiny{}$1.89_{-0.34}^{+0.38}\times10^{12}$}\tabularnewline
{\tiny{}RLQs} & {\tiny{}$1.56$} & {\tiny{}$2474$} & {\tiny{}$13.76_{-1.86}^{+1.64}$} & {\tiny{}$2.21_{-0.22}^{+0.37}$} & {\tiny{} $2.14$} & {\tiny{}$4$} & {\tiny{}$4.33\pm0.57$} & {\tiny{}$2.01_{-0.69}^{+0.84}\times10^{13}$}\tabularnewline
{\tiny{}$8.5\leq\log(M_{\textrm{BH}})\leq9.0$} & {\tiny{}$1.47$} & {\tiny{}$9446$} & {\tiny{}$8.00_{-1.28}^{+0.96}$} & {\tiny{}$1.88_{-0.15}^{+0.16}$} & {\tiny{}$0.23$} & {\tiny{}$7$} & {\tiny{}$2.51\pm0.34$} & {\tiny{}$3.98_{-1.70}^{+2.33}\times10^{12}$}\tabularnewline
{\tiny{}$9.0\leq\log(M_{\textrm{BH}})\leq9.5$} & {\tiny{}$1.47$ } & {\tiny{}$9446$} & {\tiny{}$11.39_{-0.95}^{+0.67}$} & {\tiny{}$2.60_{-0.2}^{+0.22}$} & {\tiny{}$0.63$} & {\tiny{}$6$} & {\tiny{}$3.94\pm0.32$ } & {\tiny{}$1.79_{-0.40}^{+0.46}\times10^{13}$}\tabularnewline
\hline 
 &  &  &  &  &  &  &  & \tabularnewline
\end{tabular}
\end{table*}

\end{singlespace}

\subsection{Projected correlation function $w_{p}\left(r_{p}\right)$ \label{sec:Section4-1}}

\noindent First, we check the consistency of our results by calculating
the real-space TPCF for the entire quasar sample in the interval $0.3\leq z\leq2.3$
and compare it with previous clustering studies. We select a fitting
range of $2\leq r_{p}\leq130\:h^{-1}\:\textrm{Mpc}$ to have a distance
coverage similar to previous quasar clustering studies (e.g., \citealt{2009ApJ...697.1656S}).
To determine the appropriate values for our TPCFs, we fit eq. \ref{eq:power-law}
with $r_{0}$ and $\gamma$ as free parameters using a $\chi^{2}$
minimization technique. We find a real-space correlation length of
$r_{0}=6.81_{-0.30}^{+0.29}\:h^{-1}\:\textrm{Mpc}$ and a slope of
$\gamma=2.10_{-0.05}^{+0.05}$, which is in good agreement with the
results of \citet{2009ApJ...697.1634R} for the SDSS DR5 quasar catalog,
and \citet{2010MNRAS.409.1691I} for their SDSS DR7 uniform quasar
catalog. Subsequently, we derive the best-fit $r_{0}$ and $\gamma$
values for all the quasars samples. The best-fitting values and their
respective errors are presented in Table \ref{tab:quasar_best_fitting_values}. 

\noindent We then split each redshift range according to their radio-loudness
and virial BH mass to study the clustering dependence on these properties.
The results of our clustering analysis for the different quasar sub-samples
as a function of radio-loudness are presented in the left panels of
Fig. \ref{fig:Real-space-correlations}. 

\noindent The best-fitting parameters in the interval $0.3\leq z\leq2.3$
are $r_{0}=10.95_{-1.58}^{+1.22}\:\textrm{Mpc}$, $\gamma=2.29_{-0.34}^{+0.53}$
for the RLQs and $r_{0}=6.59_{-0.24}^{+0.33}\,h^{-1}\:\textrm{Mpc},$
$\gamma=2.09_{-0.09}^{+0.10}$ for the RQQs (see Table \ref{tab:quasar_best_fitting_values}).
The latter fit is poor with $\chi^{2}=19.60$ and 7 dof, while the
former, with the same number of data points, is more acceptable, with
$\chi^{2}=1.06$ . It is clear from our clustering measurements that
RLQs are more strongly clustered than RQQs. The two additional redshift
bins show similar trends, with RLQs in the low-z bin clustering more
strongly.

\noindent In order to check our results, we estimate the correlation
function for 100 randomly selected quasar sub-samples chosen from
the RQQs with the same number of quasars as RLQs in the corresponding
redshift interval. The randomly selected quasar samples present similar
clustering lengths to those of RQQs.

We also fit the correlation function over a more restricted range
to examine the impact of different distance scales on the clustering
measurements. Using $2\leq r_{p}\leq35\:h^{-1}\:\textrm{Mpc}$, we
obtain a model with a somewhat smaller correlation scale-length $r_{0}=6.04{}_{-0.60}^{+0.51}\,h^{-1}\:\textrm{Mpc}$
and a flatter slope $\gamma=1.72{}_{-0.10}^{+0.10}$ for RQQs in the
full sample. The model matches the data better, resulting in $\chi^{2}=1.06$
and 4 dof. This may signal a change in the TPCF with scale; the transition
between the one-halo and two-halo terms may be responsible for the
$w_{p}\left(r_{p}\right)$ distortion on smaller scales (e.g., \citealt{2004MNRAS.355.1010P}).
Our remaining non-radio samples show a similar trend of improving
the fits at smaller distances. For RLQs, we obtain $\left(r_{0},\gamma\right)=\left(9.75_{-1.60}^{+1.90},\,2.70{}_{-0.60}^{+0.50}\right)$
with $\chi^{2}=2.77$ and 4 dof. The changes in the parameters are
within the error bars. 

\noindent We use the virial BH mass estimations based on single-epoch
spectra to investigate whether or not quasar clustering depends on
BH mass. The emission line which is employed to determine the fiducial
virial mass depends on the redshift interval (see \citealt{2008ApJ...680..169S}
for a description).

\noindent First, we divide the quasar samples using the median virial
BH mass in redshift intervals of $\triangle z=0.05$ following \citet{2009ApJ...697.1656S}.
Although this approach yields samples with comparable redshift distributions,
it mixes quasars regardless of their luminosity and could wash out
any true dependence on $M_{\textrm{BH}}$. Indeed, the mass samples
following this scheme hardly show any significant differences in their
clustering with correlation lengths similar to those of RQQs. Thus,
we proceed to create mass samples with two $M_{\textrm{BH}}$ intervals:
$8.5\leq\log\left(M_{\textrm{BH}}\right)\leq9.0$ and $9.0\leq\log\left(M_{\textrm{BH}}\right)\leq9.5$,
as described in Sec. \ref{sec:Section2.4-Quasarsample}. The right-hand
panels in Figure \ref{fig:Real-space-correlations} show $w_{p}(r_{p})$
for these BH mass-selected samples. It can be seen that quasars with
higher BH masses have stronger clustering. For $0.3\leq z\leq2.3$,
we obtain $r_{0}=8.535_{-2.25}^{+1.57}\,h^{-1}\:\textrm{Mpc}$, $\gamma=1.84_{-0.20}^{+0.21}$
for quasars with $8.5\leq\log\left(M_{\textrm{BH}}\right)\leq9.0$;
and $r_{0}=10.45_{-0.98}^{+0.79}\,h^{-1}\:\textrm{Mpc}$, $\gamma=2.36_{-0.17}^{+0.18}$
for BH masses in the range $9.0\leq\log\left(M_{\textrm{BH}}\right)\leq9.5$.
In the other $z-$bins, the resulting trend is similar, with the low-z
bin showing the larger clustering amplitudes. These trends hold when
the distance is restricted to $2\leq r_{p}\leq35\:h^{-1}\:\textrm{Mpc}$,
with no significant variations in $r_{0}$ and $\gamma$ due to the
larger uncertainties at these scales.

\subsection{Quasar bias factors }

\noindent 
\begin{figure}[h]
\raggedright{}\centering\includegraphics[scale=0.5,clip=true]{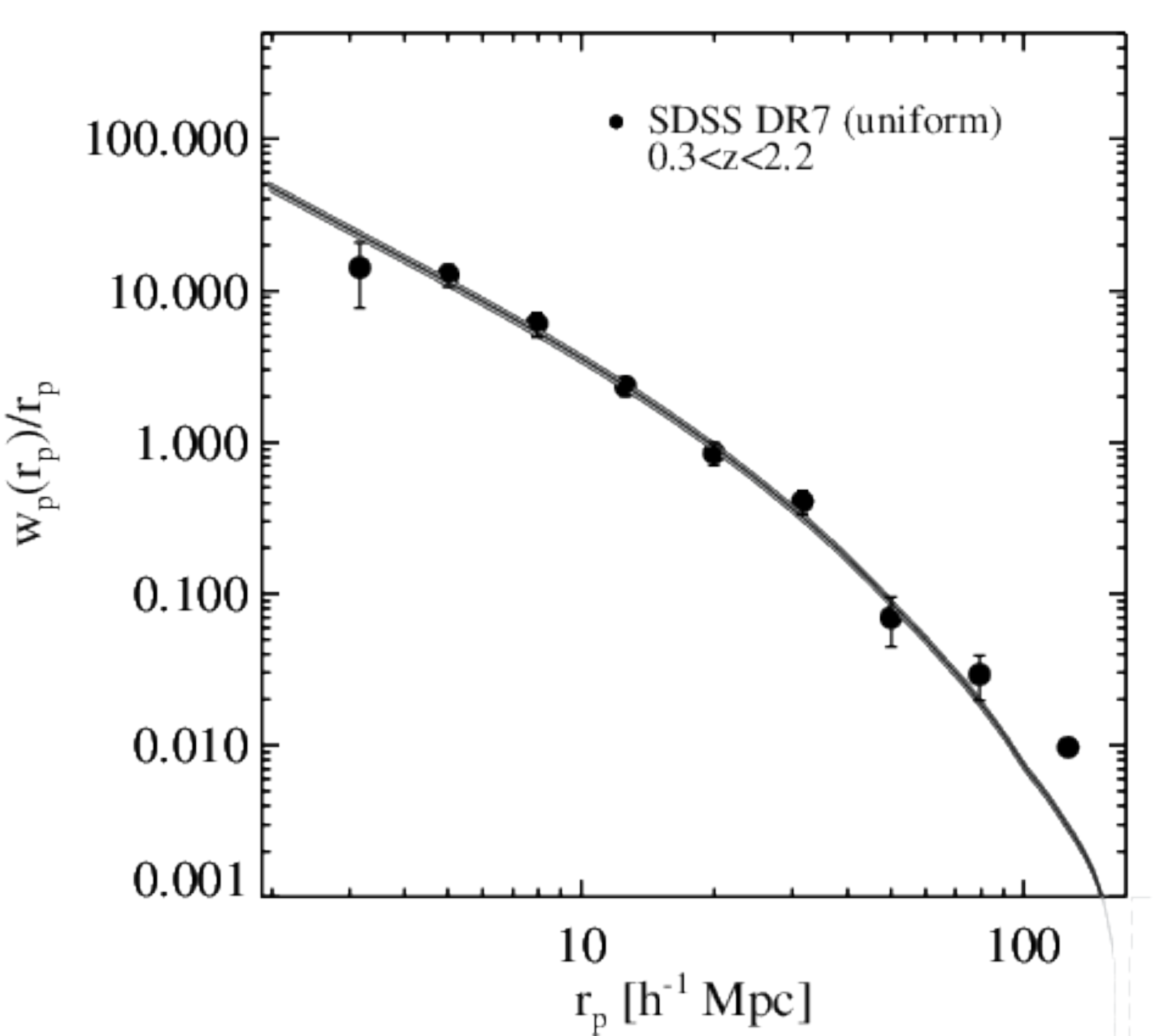}\caption{\label{fig:real-space-all}Real-space correlation function for the
SDSS DR7 quasar uniform sample with $0.3\leq z\leq2.3$. The
solid line denotes the model $w_{\textrm{QSO}}\left(r_{p}\right)$
defined in eq. \ref{eq:wdm_model} and the shaded areas are the $1-\sigma$
uncertainties.  Errors bars are the square root of the diagonal elements
from the covariance matrix computed using the jackknife method. } 
\end{figure}

\noindent 
\begin{figure*}
\centering{}\centering\includegraphics[scale=0.6]{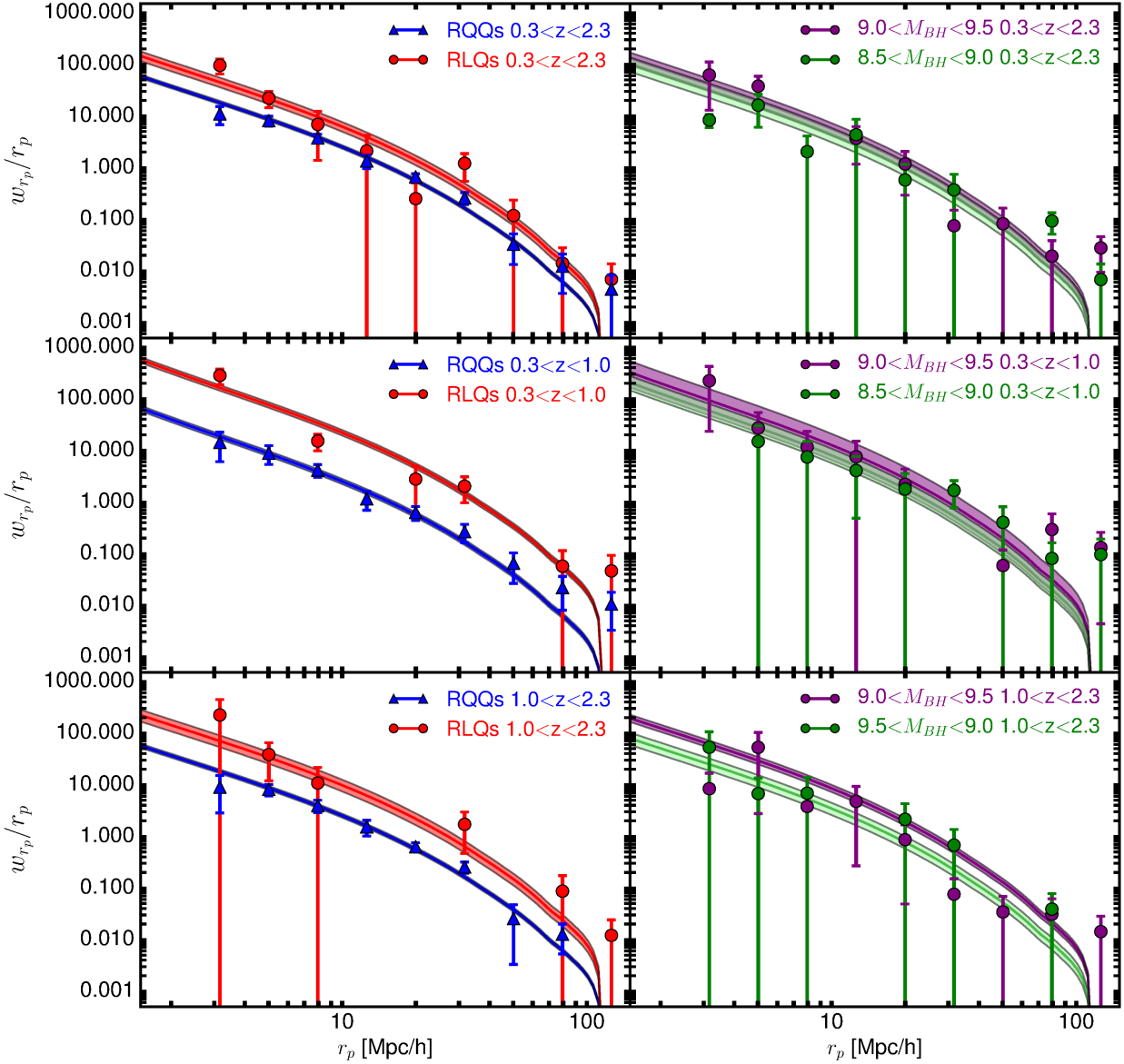}\caption{\label{fig:Real-space-correlations}Projected correlation functions
for the radio-loudness (left) and BH mass (right) samples corresponding
to the redshift intervals defined in Table \ref{tab:quasar_best_fitting_values}.
The thin lines in each panel represent the term $b^{2}\,w_{\textrm{DM}}\left(r_{p}\right)/r_{p}$
for each sample, where the shaded areas correspond to the $1-\sigma$
errors in the bias factor. }
\end{figure*}

\noindent We compute the quasar bias factors over the scales $2.0\leq r_{p}\leq130\,h^{-1}\:\textrm{Mpc}$
using the $w_{\textrm{DM}}\left(r_{p}\right)$ model in eq. \eqref{eq:wdm_model}.
Again, this distance scale has been chosen to have a good overlap
with previous SDSS quasar clustering studies (e.g., \citealt{2009ApJ...697.1656S,2009ApJ...697.1634R}).
The best-fit bias values and the corresponding typical DMH masses
for quasar samples are shown in Table \ref{tab:quasar_best_fitting_values}.
We find that the SDSS DR7 quasars at $\bar{z}=1.30$ (Figure \ref{fig:real-space-all})
have a bias of $b=2.00\pm0.08$. Previous bias estimates from 2QZ
\citep{2005MNRAS.356..415C} and 2SLAQ \citep{2008MNRAS.383..565D}
surveys are consistent with our results within the $1\sigma$ error
bars.

The left panel on Figure \ref{fig:Real-space-correlations} compares
the projected real-space TPCF $w_{p}/r_{p}$ for the RLQs (red) and
RQQs (blue). Optically selected quasars are significantly less clustered
than radio quasars in the three redshift bins analyzed, which implies
that they are less biased objects. Indeed, the RLQs and RQQs, with
mean redshifts of $\bar{z}=1.20$ and $\bar{z}=1.28$, have bias equivalent
to $b=3.14\pm0.34$ and $b=2.01\pm0.08$, respectively. These bias
factors correspond to typical DMH masses of $1.23_{-0.39}^{+0.47}\times10^{13}h^{-1}M_{\odot}$
and $2.38_{-0.38}^{+0.42}\times10^{12}\,h^{-1}M_{\odot}$, respectively.
We obtain similar results for RQQs in the other two redshift bins
with $\bar{z}=0.65$ and $\bar{z}=1.58$, respectively, (see Table
\ref{tab:quasar_best_fitting_values}). There are considerable differences
between the low-z and high-z bins results for RLQs, with low-z RLQs
residing in more massive haloes with masses of $1.16_{-0.33}^{+0.37}\times10^{14}h^{-1}M_{\odot}$.

The projected correlation functions for the mass samples are shown
in Fig. \ref{fig:Real-space-correlations} (right panels), and the
corresponding best-fit bias parameters are reported in Table \ref{tab:quasar_best_fitting_values}.
We find $b=2.64\pm0.42$ for quasars with $8.5\leq\log\left(M_{\textrm{BH}}\right)\leq9.0$,
and $b=2.99\pm0.43$ for the objects with $9.0\leq\log\left(M_{\textrm{BH}}\right)\leq9.5$
in the full redshift interval. There is a clear trend: the quasars
powered by the most massive BHs are more clustered than quasars with
less massive BHs. These quasars are more biased than RQQs, but less
than radio quasars. In the other $z-$bins, the $b$ values are comparable
to those of the full sample. This implies larger halo masses for the
low-z quasars.

We also estimate the bias over $2.0\leq r_{p}\leq35\,h^{-1}\:\textrm{Mpc}$.
RQQs in the three bins show hardly almost no difference within the
uncertainties. The resulting bias for RLQs is $b=3.11\pm0.42$ at
$0.3\leq z\leq2.2$, which is approximately $1\%$ smaller in comparison
to the bias at $2.0\leq r_{p}\leq130\,h^{-1}\:\textrm{Mpc}$. Therefore,
restricting the bias does not affect our conclusions for the radio
samples. For the mass samples, they remain virtually the same when
the range is restricted. 

\subsection{Bias and host halo mass redshift evolution}

\noindent 
\begin{figure*}
\centering{}\centering\includegraphics[scale=0.55,clip=true]{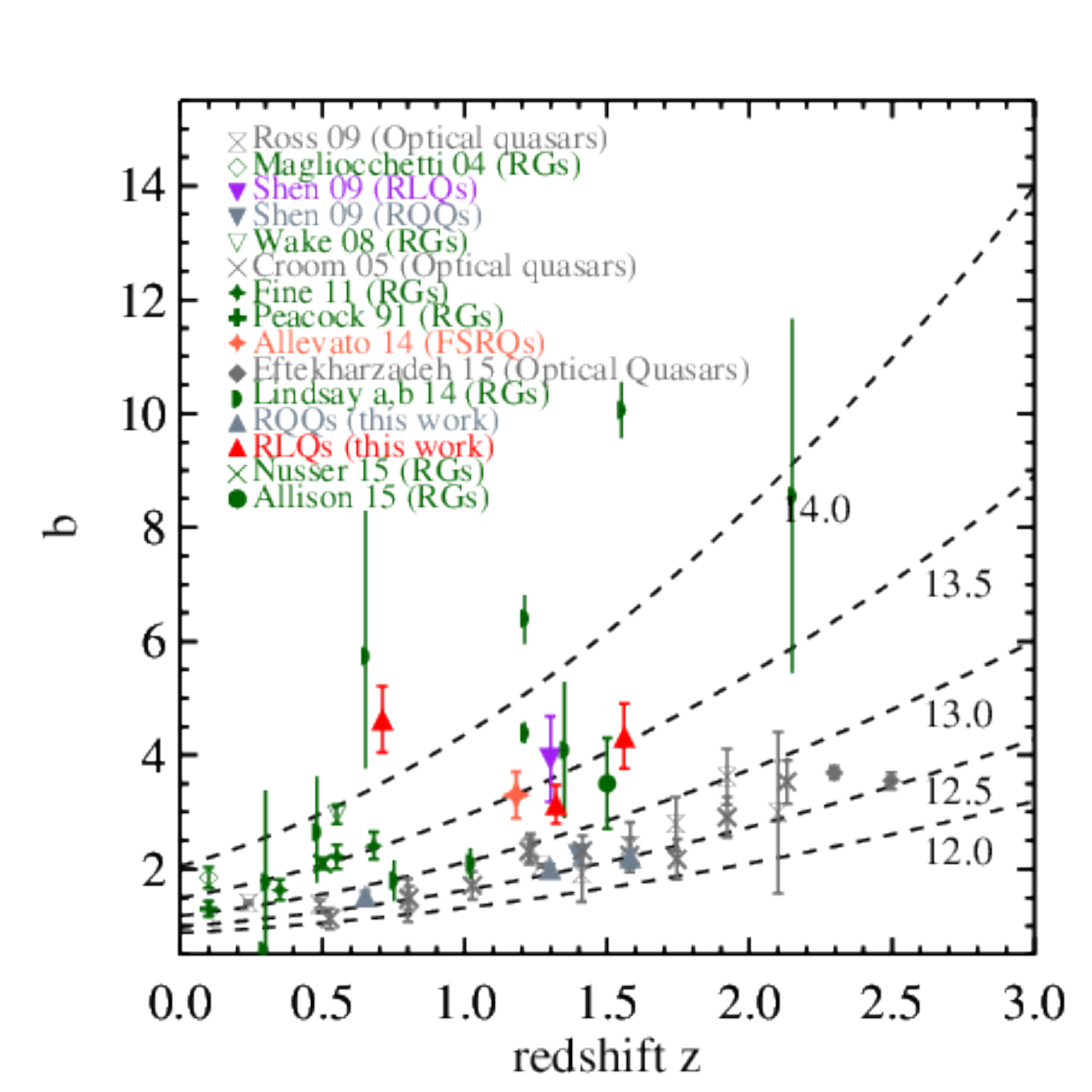}\includegraphics[scale=0.55,clip=true]{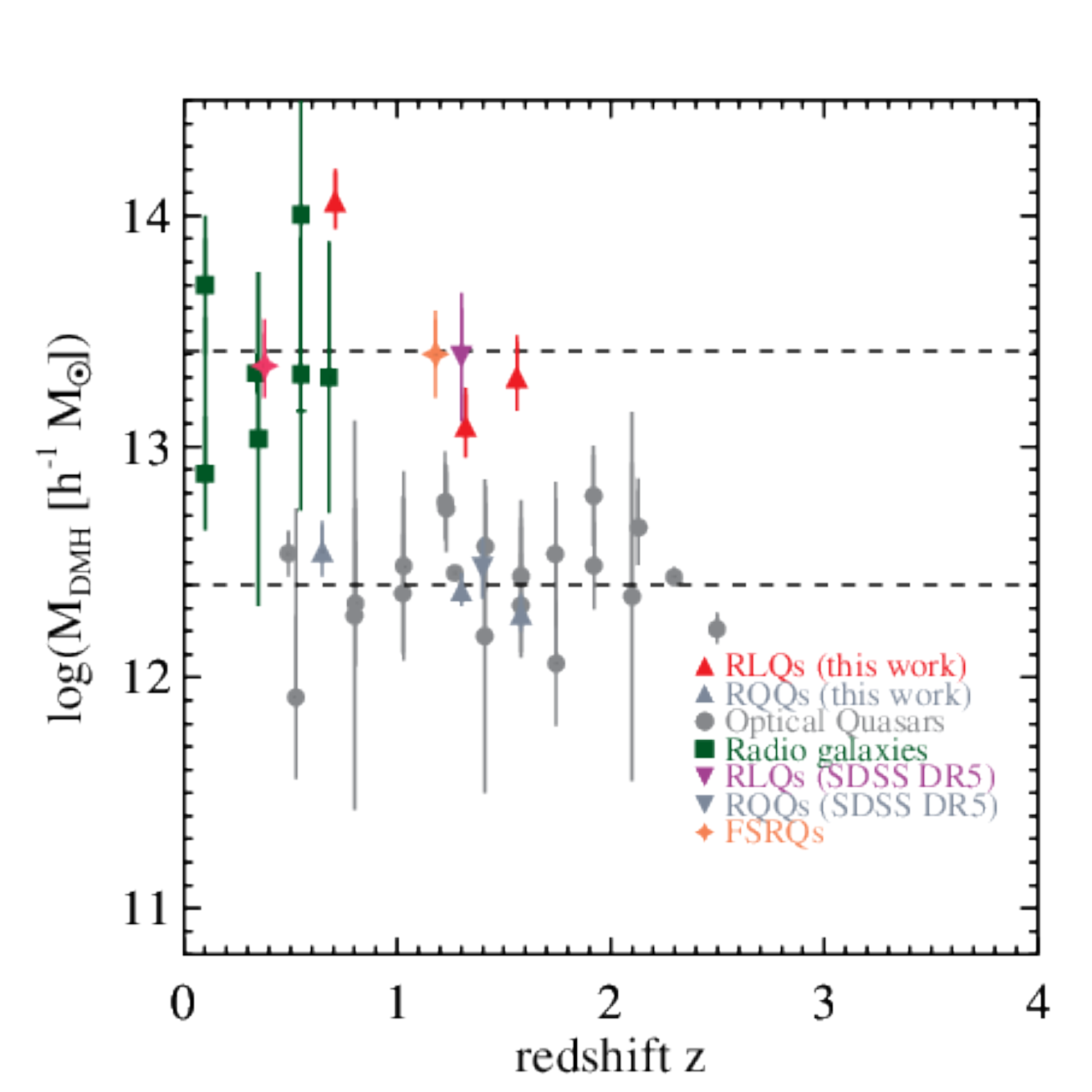}\caption{\label{fig:bias_dmh_mass}Left: The derived linear bias parameter
$b$ as a function of redshift for radio and optical AGN samples represented
by the corresponding legend. Red or gray triangles represent the RLQs
or RQQs sub-samples of this work, respectively. The dashed lines denote
the expected redshift evolution of DMH masses based on the models
from \citet{2001MNRAS.323....1S} with $\log\,\left(M_{\textrm{DM}}/h^{-1}M_{\odot}\right)=\left[12.0,\,13.0,\,13.5,\,14.0\right]$.
Right: Typical DMH masses $M_{\textrm{DMH}}$ against redshift for
RLQs and RQQs from our sample (red and gray triangles, respectively),
RLQs and RQQs from SDSS DR5 (purple and gray downward triangles, \citealt{2009ApJ...697.1656S},
respectively), optical quasars (gray circle, \citealt{2005MNRAS.356..415C,2009ApJ...697.1634R,2015MNRAS.453.2779E}),
radio galaxies (RGs, dark green squares, \citealt{1991MNRAS.253..307P,2004MNRAS.350.1485M,2008MNRAS.391.1674W,2011MNRAS.418.2251F,2014MNRAS.440.1527L,2014MNRAS.440.2322L,2015MNRAS.451..849A,2015ApJ...812...85N}),
and flat-spectrum radio quasars (FSRQs) (orange star, \citealt{2014ApJ...797...96A})
. For comparison, we show with dashed lines the mass values corresponding
to $\log\,\left(M_{\textrm{DM}}/h^{-1}M_{\odot}\right)=\left[12.4,\,13.41\right].$
When bias and mass estimations are not provided by the authors we
use the reported power-law best-fitting values to estimate $b$ and
$M_{\textrm{DMH}}$ \citep{1980lssu.book.....P,2010ApJ...713..558K}.}
\end{figure*}

In Figure \ref{fig:bias_dmh_mass} (left panel), we show our bias
estimates for RQQs and RLQs (red and gray triangles, respectively).
It can be seen that the bias is a strong function of redshift. In
the same plot, we show the previous bias estimates from the optical
spectroscopic quasar samples (gray symbols) as well as radio-loud
AGNs (green and orange symbols). Our estimates for both RQQs and RLQs
are consistent with previous works. The expected redshift evolution
tracks of DMH masses based on the models from \citet{2001MNRAS.323....1S}
are shown by dashed lines in Fig. \ref{fig:bias_dmh_mass}. RQQs follow
a track of constant mass a few times $10^{12}\,h^{-1}\,M_{\odot}$,
while the majority of RLQs and radio sources approximately follow
a track of $\sim10^{14.0-13.5}\,h^{-1}\,M_{\odot}$ within the error
bars. 

\subsection{Clustering as a function of radio-loudness}

\noindent Even though the number of radio sources is only $\sim7.6\%$
of the total number of quasars, it is clear from the left-hand panels
of Fig. \ref{fig:Real-space-correlations} that RLQs are considerably
more clustered than RQQs in all the redshift bins. The stronger clustering
presented by RLQs suggests that these inhabit more massive haloes
than their radio-quiet counterparts. The RLQs typical halo mass of
$>1\times10^{13}h^{-1}\,M_{\odot}$ is characteristic of galaxy groups
and small clusters, while the typical mass of a few times $10^{12}h^{-1}\,M_{\odot}$
for RQQs is typical of galactic haloes. The higher DMH mass presented
by RLQs in the low-z bin is similar to the halo mass of galaxy clusters,
which is usually $>1\times10^{14}h^{-1}\,M_{\odot}$.

The right-hand panel in Fig. \ref{fig:bias_dmh_mass} presents the
DMH masses against redshift for the same samples as in the left-hand
panel. Our new mass estimates for RLQs and RQQs are generally consistent
with those derived in previous works (e.g., \citealt{2005MNRAS.356..415C,2006MNRAS.371.1824P,2009ApJ...697.1634R,2009ApJ...697.1656S}).
We denote the typical halo masses for the two quasar populations using
dashed lines. This suggests that the difference between the typical
host halo masses for RLQs and RQQs is constant with redshift, with
the haloes hosting RLQs being approximately one order of magnitude
more massive. 


\subsection{Clustering as function of BH masses }

\begin{figure}[h]
\centering{}\centering\includegraphics[scale=0.46]{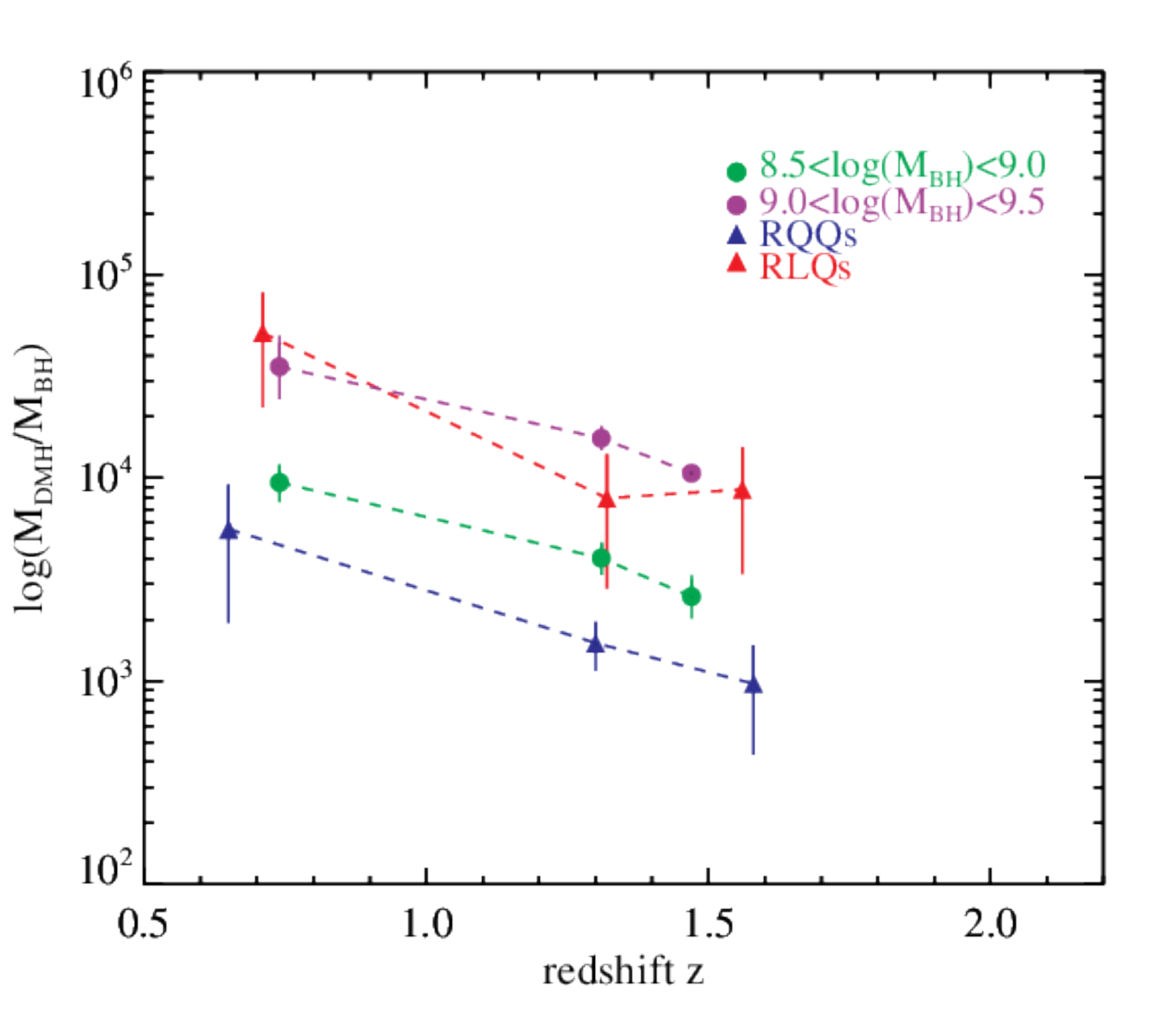}\caption{\label{fig:bhmass_vs_z} Ratio between the DMH and the average
virial BH masses for our quasar samples as a function of redshift.}
\end{figure}

Our clustering measurements for the $8.5\leq\log\left(M_{\textrm{BH}}\right)\leq9.0$
and $9.0\leq\log\left(M_{\textrm{BH}}\right)\leq9.5$ show a clear
dependence on virial BH masses. This trend is apparent in Fig. \ref{fig:Real-space-correlations}
(right panels) for all the redshift bins considered. Moreover, this
is reflected in our $M_{\textrm{BH}}$ predictions for the mass samples
in Figure \ref{fig:bhmass_vs_z}. The quasars powered by SMBHs with
$9.0\leq\log\left(M_{\textrm{BH}}\right)\leq9.5$ present larger clustering
amplitudes than those with less massive BH masses in the range $8.5\leq\log\left(M_{\textrm{BH}}\right)\leq9.0$.
Table \ref{tab:quasar_best_fitting_values} indicates that both RLQs
and the quasars with BH masses of $9.0\leq\log\left(M_{\textrm{BH}}\right)\leq9.5$
have larger correlation lengths than RQQs and quasars with $8.5\leq\log\left(M_{\textrm{BH}}\right)\leq9.0$.
However, RLQ clustering is at least slightly stronger in all the redshift
bins analyzed. It is important to remark that the use of virial estimators
to calculate the BH masses is subject to large uncertainties (e.g.,
\citealt{2008ApJ...680..169S,2012ApJ...753..125S,2012ApJ...753L...2A})
leading to significant biases and scatter around the true BH mass
values, which could potentially weaken any clustering dependence on
BH mass. Nevertheless, our results give some validity to their use
in clustering analyses.

Fig. \ref{fig:bhmass_vs_z} shows the redshift evolution of the ratio
between the DMH and the average virial BH masses for our quasar samples.
The different lines mark the ratio for each quasar sample denoted
by the plot legend. The ratios reproduce the trend for the clustering
amplitudes in all the samples: RLQs and quasars with $9.0\leq\log\left(M_{\textrm{BH}}\right)\leq9.5$
cluster more strongly than RQQs and quasars with $8.5\leq\log\left(M_{\textrm{BH}}\right)\leq9.0$,
respectively. Quasars with $9.0\leq\log\left(M_{\textrm{BH}}\right)\leq9.5$
present clustering comparable to RLQs. Also, it is evident that the
ratios are larger at low-z due to the host haloes being more massive
and the virial BH masses showing no significant changes with redshift
(see Table \ref{tab:quasar_sample}). 

An important point to consider is the cause of stronger clustering:
is the stronger clustering for the high-mass quasars due to the fact
that they are radio loud, or are the RLQs more clustered due to the
fact that they have higher BH masses. We can address this by examining
the distribution of RLQs on the virial BH mass plane. This distribution
is not restricted to high BH masses only. Instead, RLQs present BH
masses in all the ranges sampled, indicating that their radio-emission
rather than high BH mass is responsible for the stronger clustering
in RLQs. However, for the high-mass sample only a fraction of $\sim6\%$
is radio-loud, which translates to approximately $700$ RLQs, which
is not large enough to obtain a reliable clustering signal. For the
high-mass sample minus the radio-quasars, we do obtain a clustering
amplitude similar to those including radio objects. Therefore, we
conclude that the stronger clustering for both samples is mainly due
to the intrinsic properties of each sample. This point needs to be
addressed using forthcoming quasar samples with higher quasar numbers.

\subsection{Clustering as a function of redshift}

\noindent 
\begin{figure}[h]
\raggedright{}\centering\includegraphics[scale=0.55,clip=true]{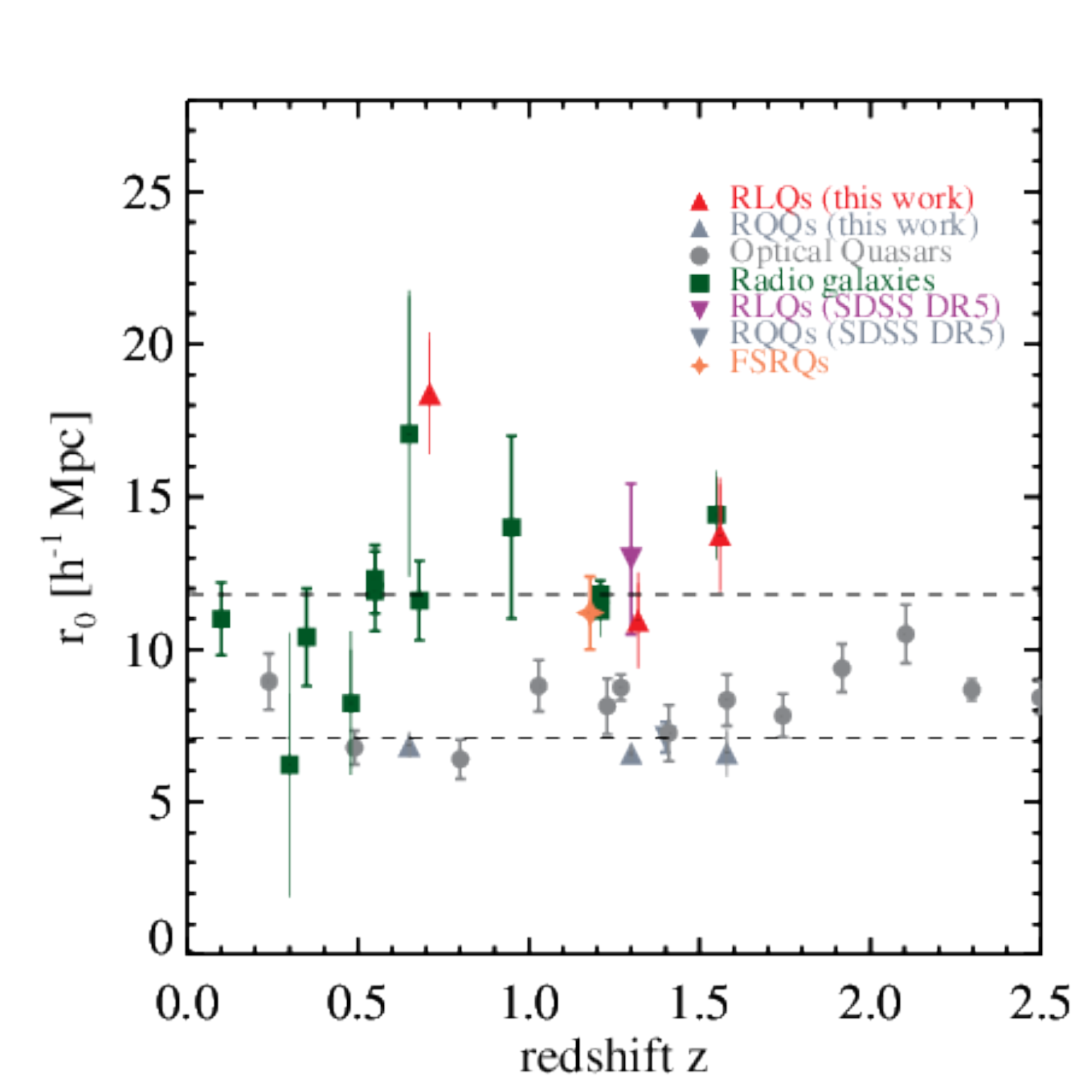}\caption{\label{fig:radio_r0} Different values for the real-space correlation
length $r_{0}$ against redshift for RLQs and RQQs from SDSS DR5 (purple
and gray downward triangles, \citealt{2009ApJ...697.1656S}), optical
quasars (gray circles, \citealt{2005MNRAS.356..415C,2009ApJ...697.1634R,2015MNRAS.453.2779E}),
radio galaxies (dark green squares, \citealt{1991MNRAS.253..307P,2004MNRAS.350.1485M,2008MNRAS.391.1674W,2011MNRAS.418.2251F,2014MNRAS.440.1527L}),
and FSRQs (orange star, \citealt{2014ApJ...797...96A}). The $r_{0}$
values for RLQ and RRQ in our sample are represented by red and gray
upward triangles, respectively. For comparison, we show the $r_{0}$
values corresponding to $r_{0}=\left[11.8,\,7.1\right]\,h^{-1}\:\textrm{Mpc}$
(dashed lines). The results from \citet{2014MNRAS.440.1527L} are
derived assuming linear clustering.}
\end{figure}

In Fig. \ref{fig:radio_r0}, we show our $r_{0}$ measurements along
with results from previous works for radio galaxies \citep{1991MNRAS.253..307P,2004MNRAS.350.1485M,2008MNRAS.391.1674W,2011MNRAS.418.2251F,2014MNRAS.440.1527L,2015MNRAS.451..849A,2015ApJ...812...85N},
optically-selected quasars \citep{2009ApJ...697.1634R,2005MNRAS.356..415C,2015MNRAS.453.2779E},
and $\gamma-\textrm{selected}$ blazars \citep{2014ApJ...797...96A}.
In these samples, the typical $1.4\:\textrm{GHz}$ radio-luminosities
for AGNs is $10^{23}$-$10^{26}\,\textrm{W}\,\textrm{Hz}^{-1}$ which
is representative of FRI sources, whilst for our sample the average
radio-luminosity is $\sim8\times10^{26}\,\textrm{W}\,\textrm{Hz}^{-1}$,
which is near the boundary between FRI and FRII sources. 

A systematic trend with redshift is observed in Fig. \ref{fig:radio_r0},
which indicates that the majority of radio sources considered have
clustering lengths over the entire redshift range considered $(0<z<2.3)$.
This is consistent with the trend from Fig. \ref{fig:bias_dmh_mass},
where the majority of radio sources seem to inhabit haloes of $M_{\textrm{DMH}}>1\times10^{13}\,$
at all redshifts. The simplest interpretation of this result is that
a considerable part of the bright radio population resides in massive
haloes with large correlation lengths. Our new RLQ clustering measurements
for the full sample and high-z bin agree, within the errors bars,
with the previous single estimation from \citet{2009ApJ...697.1656S}
using the SDSS DR5 quasar sample, while the low-z bin correlation
amplitude is consistent with \citet{2014MNRAS.440.1527L}.

\citet{2003AA...405...53O} measured the angular TPCF for the NVSS
survey \citep{1998AJ....115.1693C} and concluded that lower luminosity
radio sources ($\leq10^{26}\,\textrm{W}\,\textrm{Hz}^{-1}$) present
typical correlation lengths of $r_{0}\lesssim6\,h^{-1}\:\textrm{Mpc}$,
whilst the brighter radio sources ($>10^{26}\,\textrm{W}\,\textrm{Hz}^{-1}$),
mainly FRII type, have significantly larger scale lengths of $r_{0}\gtrsim14\,h^{-1}\:\textrm{Mpc}$.
Our findings are consistent with \citet{2003AA...405...53O} predictions
for the bright radio population. It is possible that the weaker correlation
length presented by lower radio-luminosity samples in Fig. \ref{fig:radio_r0}
indicates a mild clustering dependence on radio-luminosity. However,
our RLQs sample is still too small to draw firm conclusions on the
radio luminosity dependence as the increasing errors for these luminosity-limited
samples mean we cannot satisfactorily distinguish between them

\noindent The DMH masses for RLQs and quasars with $9.0\leq\log\left(M_{\textrm{BH}}\right)\leq9.5$
at $0.3<z<1.0$, are approximately $>1\times10^{14}h^{-1}\,M_{\odot}$,
which is the typical value for cluster-size haloes. Moreover, these
halo masses are larger than the corresponding haloes for quasar samples
at $z>1.0$. This suggests that the environments in which these objects
reside is different from those of their high-z counterparts. Additionally,
the radio source clustering amplitudes are similar to the clustering
scale of massive galaxy clusters (e.g., \citealt{2003ApJ...599..814B}).
This almost certainly reveals a connection between quasar radio-emission
and galaxy cluster formation that must be explored in detail with
data from forthcoming radio surveys.

\subsection{Clustering and AGN unification theories}

Our clustering results hint at an interesting point regarding the
relationship between RLQs and radio galaxies in AGN classifications,
which consider these AGNs as the same source type seen from different
angles (e.g., \citealt{1995PASP..107..803U}). Thus, we would expect
that different AGN types such as radio galaxies and RLQs, should have
similar clustering properties. The real-space correlation lengths
for RLQs (red triangles) and other radio sources including, radio
galaxies (green squares), are shown in Fig. \ref{fig:radio_r0}. We
see that there is a reasonable consistency for most $r_{0}$ values
up to $z\lesssim2.3$. We identify the same trend in Fig. \ref{fig:bias_dmh_mass}
(right panel), where bright radio sources seem to inhabit haloes of
approximately constant mass of $\gtrsim10^{13.5}\,h^{-1}\,M_{\odot}$.
Our clustering study seems to support the validity of unification
models at least for RLQs and radio galaxies with relatively median
radio-luminosities ($\gtrsim1\times10^{23}\,\textrm{W}\,\textrm{Hz}^{-1}$). 

\citet{2014ApJ...797...96A} studied the clustering properties of
a $\gamma-\textrm{selected}$ sample of blazars divided into BL Lacs
and flat-spectrum radio quasars (FSRQs). In the context of unification
models, FSRQs are associated with intrinsically powerful FRII radio
galaxies, while BL Lacs are related to weak FRI radio galaxies. From
a clustering point of view, as explained before, luminous blazars
should have similar clustering properties to radio galaxies. In Figs.
\ref{fig:bias_dmh_mass} and \ref{fig:radio_r0}, we denote by a orange
star, the DMH mass and correlation length for FSRQs, respectively,
found by \citet{2014ApJ...796....4A}. FSRQs show a similar $M_{\textrm{DMH}}$
value to those of radio galaxies and RLQs, supporting a scenario in
which radio AGNs such as quasars, radio galaxies and powerful blazars
are similar from a clustering perspective and reside in massive hosting
haloes providing the ideal place to fuel the most massive and powerful
BHs. 

Based on an analysis of the cross-correlation function for radio galaxies,
RLQs and a reference sample of luminous red galaxies \citet{2010MNRAS.407.1078D}
concluded that the clustering for RLQs is weaker in comparison with
radio galaxies. This is apparently at odds with previous clustering
measurements and our results. However, there are several differences
between Donoso's and our sample that must be considered. First, Donoso's
sample is significantly smaller with only 307 RLQs at $0.35<z<0.78$.
Secondly, in the common range between the two samples where the TPCF
is computed, their clustering signal has large uncertainties. Thirdly,
they compute the clustering for objects with radio-luminosities restricted
to $>10^{25}\,\textrm{W}\,\textrm{Hz}^{-1}$. We employ the same luminosity
cut only for the high-z bin, while for the low-z bin only sources
brighter than $>4\times10^{24}\,\textrm{W}\,\textrm{Hz}^{-1}$ are
considered. The mean luminosity for both redshift bins is $>2\times10^{26}\,\textrm{W}\,\textrm{Hz}^{-1}$
(see Table \ref{tab:quasar_sample}). Therefore, comparable radio-luminosity
cuts were used for both samples. For these reasons, it is difficult
to draw any conclusions from comparison with the Donoso results.

\subsection{The role of mergers in quasar radio-activity }

\noindent 
\begin{figure*}
\centering{}\centering\includegraphics[scale=0.61]{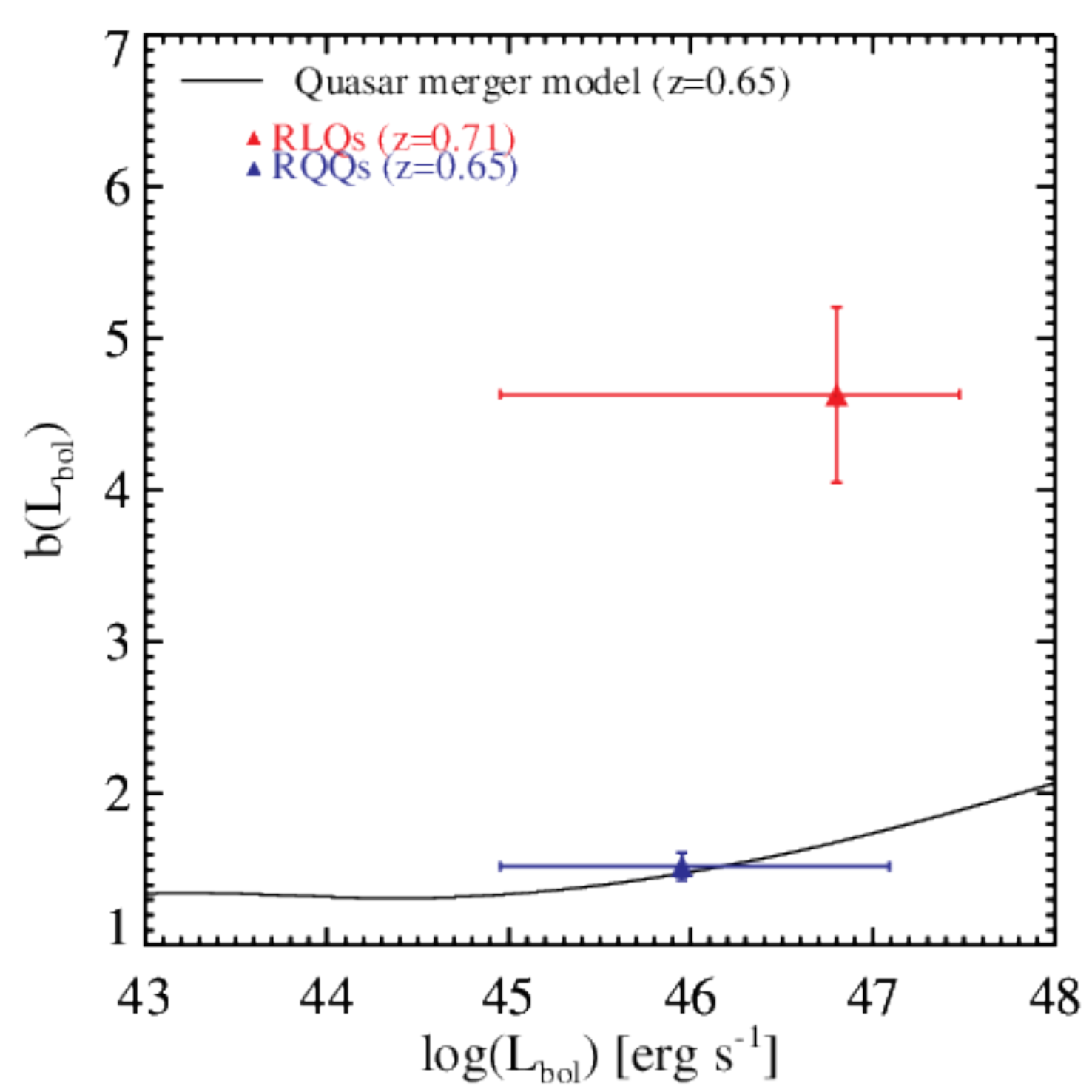}\includegraphics[scale=0.61]{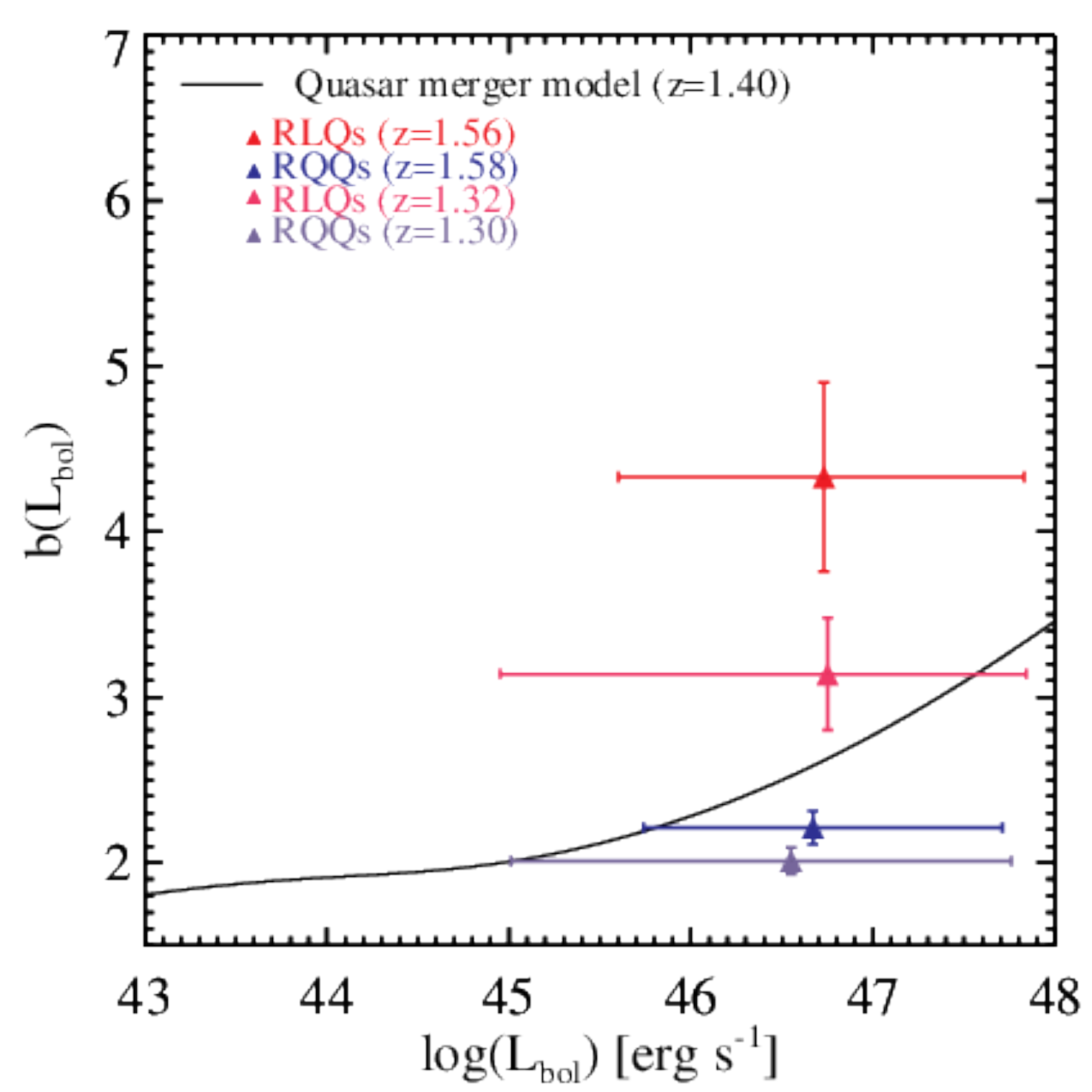}\caption{\label{fig:BH_accretion_models} Bias parameter $b$ as a function
of bolometric luminosity for our RLQs and RQQs in the ranges $0.3\leq z\leq1.0$
(left) and $1.0\leq z\leq2.0$ (right). Errors in the $\textrm{L}_{bol}$
axis are the dispersion values for each different quasar sample. The
solid lines in both panels denote the predicted bias luminosity evolution
according to the \citet{2009ApJ...704...89S} model, which predicts
that quasar activity is triggered by galaxy mergers. }
\end{figure*}


We compare our clustering measurements with the theoretical framework
for the growth and evolution of SMBHs introduced by \citet{2009ApJ...704...89S}.
This model links the quasar properties and host halo mass with quasar
activity being triggered by major galaxy mergers. The bias factor
is a function of the instantaneous luminosity and redshift, with most
luminous quasars having larger host-halo masses. The rate of quasar
activity is controlled by the fraction parameter $f_{\textrm{QSO}}$,
which involves exponential cutoffs at both high and low mass ends
assigned according to phenomenological rules. At low masses, the cutoffs
prevent quasar activity on the smallest postmerger haloes, while those
at the highest masses cause that gas accretion to become inefficient
and subsequent BH growth stops. Figure \ref{fig:BH_accretion_models}
presents the predicted linear bias as a function of bolometric luminosity
at $z=0.65$ (left) and $z=1.40$ (right). In the low-z bin $(0.3\leq z\leq1.0)$,
the model can reproduce the bias for the RQQs. However, the quasar
merger model disagrees with the higher bias value for RLQs. At high-z
$(1.0\leq z\leq2.3)$, the consistency between the model predictions
and the measured bias for RQQs for the high-z bin and the complete
quasar sample worsens. The bias luminosity-dependent trend predicted
by the model seems to be followed slightly better by the RLQs than
in the low-z bin. 

The discrepancy between the merger-driven model predictions and our
bias values might indicate differences in the fueling channels for
both quasar types. First, our bias estimates for RQQs in the context
of the \citet{2009ApJ...697.1656S} framework favor accretion of cold
gas via galaxy mergers (referred to as cold-gas accretion). These
$M_{\textrm{DH}}$ masses are in agreement with the halo mass-scale
of a few times $\gtrsim10^{12}\,h^{-1}\,M_{\odot}$ predicted by merger-driven
models for optical quasars (e.g., \citealt{2005MNRAS.356..415C,2009ApJ...697.1634R}).
In contrast, the bias results for RLQs, which correspond to halo masses
of $\gtrsim10^{13}\,h^{-1}\,M_{\odot}$, cannot be reproduced by models
that assume that quasar activity is solely triggered by typical galaxy
mergers.

A similar difference in DMH masses has been reported in clustering
studies for X-ray selected AGNs with moderate luminosity $\left(\textrm{L}_{bol}\sim10^{43-46}\,\textrm{erg}\,\textrm{s}^{-1}\right)$
\citep{2005AA...430..811G,2009AA...494...33G,2011ApJ...741...15S,2011ApJ...736...99A,2013MNRAS.430..661M,2012MNRAS.420..514M}.
The DMH masses of X-Ray AGNs are approximately $10^{13}\,h^{-1}\,M_{\odot}$,
which is significantly higher in comparison with relatively bright
optical quasars $\left(\textrm{L}_{bol}\gtrsim10^{46}\,\textrm{erg}\,\textrm{s}^{-1}\right)$
with $\gtrsim10^{12}\,h^{-1}\,M_{\odot}$ \citep{2005MNRAS.356..415C,2009ApJ...697.1634R}.
Several authors have observationally \citep{2011ApJ...736...99A,2012MNRAS.420..514M,2014ApJ...796....4A}
and theoretically \citep{2012MNRAS.419.2797F,2013MNRAS.435..679F}
interpreted these two mass scales as evidence favoring different accretion
channels for each AGN population. \citet{2013MNRAS.435..679F}, using
semi-analytical galaxy formation models, found that cold gas fuelling
cannot reproduce the DMH masses from X-Ray AGN clustering studies.
Instead, they found that when gas cooled from quasi-hydrostatic hot-gas
haloes (i.e., known as hot-mode; \citealt{2006MNRAS.365...11C}) is
included, a much better agreement with the DMH masses derived from
X-Ray AGN clustering studies is obtained. 

\noindent The differences in DMH masses for X-Ray AGNs and optical
quasars is reminiscent of our results for RQQs and RLQs. This may
suggest that the contribution of hot-gas accretion increases for more
massive haloes, such as those hosting X-Ray AGNs and RLQs. However,
this scenario for RLQs still needs to be confronted with more detailed
simulations and models to further constrain the physics of BH accretion. 


\subsection{Black hole properties involved in quasar triggering }

As considering only cold accretion via mergers cannot explain the
mass scales associated with RQQs and RLQs, it is important to take
into account different mechanisms related to quasar activity. For
instance, the massive haloes where these RLQs are embedded must have
an important role in determining  the BH properties and the onset
of radio activity. Indeed, the BH spin could be altered by environmental
conditions: either by means of coherent gas accretion, or by BH-BH
mergers. In the spin paradigm proposed by \citet{1995ApJ...438...62W},
the rapidly spinning BHs are associated with radio-loud AGNs, whilst
the slower spinning ones are considered to be radio-quiet. Objects
above a certain spin threshold could have the necessary energy to
produce powerful relativistic jets \citep{1977MNRAS.179..433B}. The
intrinsic scatter on the BH spin values required to power the jets
may reproduce the different morphologies and the shape of the luminosity
function at radio wavelengths \citep{2011MNRAS.410...53F}. Another
plausible scenario is a two-way interaction between RLQs jets and
the surrounding intergalactic medium, as suggested by the morphological
associations of radio continuum with extended optical emission \citep{1985ApJ...290..496V},
and bent radio structures in nearby radio active galaxies \citep{1986ApJ...301..841O}.
As radio jets propagate into a dense interstellar medium they suffer
from both depolarization and decollimation that yield an enhancement
in their radio brightness \citep{1984RvMP...56..255B}. The luminosity
boosting for these objects may help to make them just bright enough
to be detectable above the FIRST survey flux limit. Finally, the magnetic
field configurations derived from polarimetry studies (e.g., \citealt{1984ARAA..22..319B})
indicate that the magnetic field in FR-II radio-galaxies is predominantly
aligned along the jet for most of its length, whereas FR-I objects
are characterized by perpendicular and parallel components. This may
suggest a correlation between the DMH mass and the efficiency in producing
the magnetic field alignment required to produce brighter radio emission. 

\noindent In conclusion, the interplay between all these BH properties
in triggering radio activity is still poorly understood. Additional
observational and theoretical efforts are required to obtain a better
comprehension of the origins of radio-emission in quasars.

\section{Summary\label{sec:Section8}}

In this study, we have investigated the quasar clustering dependence
on radio-loudness and BH virial mass, by using a sample of approximately
$48000$ spectroscopically confirmed quasars at $0.3\leq z\leq2.3$
drawn from SDSS DR7 quasar catalog \citep{2011ApJS..194...45S,2010AJ....139.2360S}.
Our radio sample consists of FIRST-detected quasars. The main conclusions
of this paper are the following:
\begin{enumerate}
\item We studied the spatial clustering of quasars at $0.3\leq z\leq2.3$
over the scales $2.0\leq r_{0}\leq130\,h^{-1}\:\textrm{Mpc}$. For
RQQs, we find a real-space correlation length equal to $r_{0}=6.59_{-0.24}^{+0.33}\,h^{-1}\:\textrm{Mpc}$
with a slope of $\gamma=2.09_{-0.09}^{+0.10}$. RLQs are more strongly
clustered than RQQs with $r_{0}=10.95_{-1.58}^{+1.22}\,h^{-1}\:\textrm{Mpc}$,
$\gamma=2.29_{-0.34}^{+0.53}$. 
\item We estimated the linear bias for RQQs and RLQs by splitting the quasar
sample according to radio-loudness, and find $b=2.01\pm0.08$ and
$b=3.14\pm0.34$, respectively, for the full redshift interval.
\item We investigated the clustering dependency on BH virial mass using
quasar samples with $8.5\leq\log\left(M_{\textrm{BH}}\right)\leq9.0$
and $9.0\leq\log\left(M_{\textrm{BH}}\right)\leq9.5$ constructed
to have comparable optical luminosity distributions. We find a dependence
on BH mass, with the quasars powered by the most massive BHs having
larger correlation lengths. These results suggest that BH virial mass
estimations based on broad emission lines may be valid BH mass proxies
for clustering studies. 
\item Using our best-fit bias values, we find that RLQs in our sample inhabit
massive haloes with masses of $M_{\textrm{DMH}}\gtrsim10^{13.5}\,h^{-1}\,M_{\odot}$
at all redshifts, which corresponds to the mass scale of galaxy groups
and galaxy clusters. RQQs reside in less massive haloes of a few times
$\sim10^{12}\,h^{-1}\,M_{\odot}$. 
\item RQQs have smaller DMH masses in comparison with RLQs. The BH mass
selected samples have larger DMH masses than RQQs, but smaller DMH
masses than those of radio quasars. However, RLQs have the most massive
DMHs in all the redshift bins considered. We considered the ratio
between the DHM and average virial BH masses for all the samples.
The ratios present the same above-mentioned trends.
\item Within our quasar sample, we do detect significant correlations between
quasar clustering and redshift for RLQs up to $z\lesssim2.3$. At
low-z, RLQs and quasars with $9.0\leq\log\left(M_{\textrm{BH}}\right)\leq9.5$
have clustering amplitudes of $r_{0}\gtrsim18\,h^{-1}\:\textrm{Mpc}$,
comparable to those of today's massive galaxy clusters. Our real-space
clustering length $r_{0}$ estimate for the full samples agrees very
well with the majority of previous complementary and independent clustering
estimates for radio galaxies and RLQs. 
\item We used radio-loudness to separate the quasar sample into RLQs and
RQQs. Our clustering measurements suggest that there are differences
between RLQs and RQQs in terms of halo and BH mass scales. Our result
is consistent with the hierarchical clustering scenario, in which
most massive galaxies harboring the most massive BHs form in the highest
density peaks, thus cluster more strongly than less massive galaxies
in typical peaks. This is confirmed by clustering analysis of the
mass samples and their dependence on $M_{\textrm{BH}}$.
\item Comparing our linear bias and DMH mass estimates with the theoretical
predictions of the merger-driven model from \citet{2009ApJ...704...89S},
we find that this model cannot explain the larger bias and DHM masses
for RLQs, suggesting that cold accretion driven by galaxy mergers
is unlikely to be the main fueling channel for RLQs with $M_{\textrm{DMH}}\geqslant10^{13}\,h^{-1}\,M_{\odot}$.
Conversely, merger model predictions agree well with our bias and
host mass estimates for RQQs, with $M_{\textrm{DMH}}\gtrsim10^{12}\,h^{-1}\,M_{\odot}$. 
\item The disagreement between the bias luminosity-dependent trend predicted
by the \citet{2009ApJ...704...89S} merger model and our bias estimates
for RLQs suggests a scenario where the radio emission is a complex
phenomenon that may depend on several BH properties such as: BH spin,
environment, magnetic field configuration, and accretion physics.
\item The similarity in clustering amplitude and host halo masses for radio-galaxies,
radio-selected AGNs, RLQs, and FSRQs is in line with the idea that
the different spectral features for these radio sources depend only
on the orientation angle and not on the environment in which they
are embedded, supporting orientation-driven unification models (\citealt{1995PASP..107..803U}
and references therein). \citet{2010MNRAS.407.1078D} found that the
clustering properties for RLQs and radio galaxies differ, with the
latter displaying a stronger clustering. In principle, these results
are in tension with our results and previous clustering studies of
radio sources (e.g. \citealt{2002MNRAS.333..100M,2008MNRAS.391.1674W,2009ApJ...697.1656S}).
However, their small sample size and large uncertainties in the clustering
in comparison with our sample make it difficult to draw any significant
conclusions. In future studies, larger samples of quasars and radio
galaxies may provide new information about the clustering properties
for both populations. 
\end{enumerate}
\begin{acknowledgements}

ERM wish to thank F. Mernier and E. Rigby for critical reading and F. Shankar for providing us the tracks in Fig. 13. 
ERM acknowledges financial support from NWO Top project, No. 614.001.006. HR acknowledges support from the ERC Advanced Investigator program NewClusters 321271. Moreover, we would like to thank the referee for valuable suggestions on the manuscript. \\

Funding for the SDSS and SDSS-II has been provided by the Alfred P. Sloan Foundation, the Participating Institutions, the National Science Foundation, the U.S. Department of Energy, the National Aeronautics and Space Administration, the Japanese Monbukagakusho, the Max Planck Society, and the Higher Education Funding Council for England. The SDSS Web Site is http://www.sdss.org/. The SDSS is managed by the Astrophysical Research Consortium for the Participating Institutions. The Participating Institutions are the American Museum of Natural History, Astrophysical Institute Potsdam, University of Basel, University of Cambridge, Case Western Reserve University, University of Chicago, Drexel University, Fermilab, the Institute for Advanced Study, the Japan Participation Group, Johns Hopkins University, the Joint Institute for Nuclear Astrophysics, the Kavli Institute for Particle Astrophysics and Cosmology, the Korean Scientist Group, the Chinese Academy of Sciences (LAMOST), Los Alamos National Laboratory, the Max-Planck-Institute for Astronomy (MPIA), the Max-Planck-Institute for Astrophysics (MPA), New Mexico State University, Ohio State University, University of Pittsburgh, University of Portsmouth, Princeton University, the United States Naval Observatory, and the University of Washington.

\end{acknowledgements}

\bibliographystyle{aa}
\bibliography{my_bib}

\end{document}